\documentclass[iop]{emulateapj}

\shorttitle{Resolving the ISM in $z \sim 2.5$ ALESS SMGs}
\shortauthors{Calistro Rivera et al.}

\newcommand{\kms}{\,km\,s$^{-1}$} 
\newcommand{\msun}{M$_{\odot}$} 

\usepackage[T1]{fontenc}
\usepackage{ae,aecompl}
\usepackage{newtxtext,newtxmath}

\begin{document}

\title{Resolving the ISM at the peak of cosmic star formation with ALMA - \\The distribution of CO and dust continuum in $\lowercase{z} \sim 2.5$ sub-millimetre galaxies}

\author{Gabriela Calistro Rivera \altaffilmark{1}}
\author{J. A. Hodge\altaffilmark{1}}
\author{Ian Smail\altaffilmark{2}}
\author{A. M. Swinbank\altaffilmark{2}}
\author{A. Wei\ss\altaffilmark{3}}
 \author{J. L. Wardlow\altaffilmark{2}}
\author{ F. Walter\altaffilmark{4}}
\author{ M. Rybak\altaffilmark{1}}
\author{ Chian-Chou Chen\altaffilmark{5}}
\author{ W. N. Brandt\altaffilmark{7,8,9}}
\author{ K. Coppin\altaffilmark{10}}
\author{E. da Cunha\altaffilmark{11}}
\author{H. Dannerbauer\altaffilmark{12, 13}}
\author{ T. R. Greve\altaffilmark{14}}
\author{A. Karim\altaffilmark{15}}
\author{ K. K. Knudsen\altaffilmark{16}}
\author{E. Schinnerer\altaffilmark{4}}
\author{J. M. Simpson\altaffilmark{6}}
\author{B. Venemans\altaffilmark{4}}
\author{P. P. van der Werf\altaffilmark{1}}

\altaffiltext{1}{Max Planck Institute for Astronomy, K\"onigstuhl 17, D-69117 Heidelberg, Germany}

\altaffiltext{1}{Leiden Observatory, Leiden University, P.O. Box 9513, 2300 RA Leiden, The Netherlands}
\altaffiltext{2}{Centre for Extragalactic Astronomy,Department of Physics,Durham University, South Road, Durham DH1 3LE, UK}
\altaffiltext{3} {Max-Planck-Institut f\"ur Radioastronomie, Auf dem H\"ugel 69, D-53121 Bonn, Germany}
\altaffiltext{4}{ Max--Planck Institut f\"ur Astronomie, K\"onigstuhl 17, 69117 Heidelberg, Germany}
\altaffiltext{5} {European Southern Observatory, Karl Schwarzschild Strasse 2, Garching, Germany}
\altaffiltext{6} {Academia Sinica Institute of Astronomy and Astrophysics, No. 1, Sec. 4, Roosevelt Rd., Taipei 10617, Taiwan}
\altaffiltext{7} {Department of Astronomy \& Astrophysics, 525 Davey Lab, Pennsylvania State University, University Park, PA 16802, USA}
\altaffiltext{8} {Institute for Gravitation and the Cosmos, Pennsylvania State University, University Park, PA 16802, USA}
\altaffiltext{9} {Department of Physics, The Pennsylvania State University, University Park, PA 16802, USA}
\altaffiltext{10} {Centre for Astrophysics Research, University of Hertfordshire, Hatfield, AL10 9AB, UK}
\altaffiltext{11} {The Australian National University, Mt Stromlo Observatory, Cotter Rd, Weston Creek, ACT 2611, Australia }
\altaffiltext{12} {Instituto de Astrof\'isica de Canarias (IAC), E-38205 La Laguna, Tenerife, Spain}
\altaffiltext{13} {Universidad de La Laguna, Dpto. Astrofisica, E-38206 La
Laguna, Tenerife, Spain}
\altaffiltext{14} {Department of Physics and Astronomy, University College London, Gower Street, London, WC1E 6BT, UK}
\altaffiltext{15}{Argelander--Institute of Astronomy, Bonn University, Auf dem H\"ugel 71, D-53121 Bonn, Germany}
\altaffiltext{16} {Department of Earth and Space Sciences, Chalmers University of Technology, Onsala Space Observatory, 439 92 Onsala, Sweden}

\email{calistro@strw.leidenuniv.nl}

\begin{abstract}
We use ALMA observations of four sub-millimetre galaxies (SMGs) at $z\sim2-3$ to investigate the spatially resolved properties of the inter-stellar medium (ISM) at scales of 1--5 kpc (0.1--0.6$''$).
The velocity fields of our sources, traced by the $^{12}$CO($J$=3-2) emission, are consistent with disk rotation to first order, implying average dynamical masses of $\sim$3$\times10^{11}$\msun~within two half-light radii.
Through a Bayesian approach we investigate the uncertainties inherent to dynamically constraining total gas masses. We explore the covariance between the stellar mass-to-light ratio and CO-to-H$_{2}$ conversion factor, $\alpha_{\rm CO}$, finding values of $\alpha_{\rm CO}=1.1^{+0.8}_{-0.7}$ for dark matter fractions of 15\%. 
We show that the resolved spatial distribution of the gas and dust continuum can be uncorrelated to the stellar emission, challenging energy balance assumptions in global SED fitting.
Through a stacking analysis of the resolved radial profiles of the CO(3-2), stellar and dust continuum emission in SMG samples, we find that the cool molecular gas emission in these sources (radii $\sim$5--14 kpc) is clearly more extended than the rest-frame $\sim$250 $\mu$m dust continuum by a factor $>2$.
We propose that assuming a constant dust-to-gas ratio, this apparent difference in sizes can be explained by temperature and optical-depth gradients alone.
Our results suggest that caution must be exercised when extrapolating morphological properties of dust continuum observations to conclusions about the molecular gas phase of the ISM.
\end{abstract}
\begin{keywords}
{galaxies: sub-millimeter, etc.}
\end{keywords}



\begingroup
\let\clearpage\relax
\endgroup
\newpage

\section{Introduction}

The process of massive galaxy assembly in the Universe has been identified to peak between $1<z<3$, where the majority of the stars in the present-day galaxies formed \citep[e.g., ][]{madau14}.
This star formation (SF) is likely strongly connected to the gas content and its distribution in the interstellar medium (ISM) and the efficiency with which this gas is transformed into stars \citep[e.g.][]{decarli16, scoville17, tacconi17}.
Gas-rich, dusty galaxies, as submillimetre galaxies \citep[SMGs; e.g.,  ][]{blain02} are effective laboratories to characterize this star-forming ISM due to their high molecular gas content \citep{bothwell13}
and their bright dust continuum emission ensured by their selection.
Moreover, they appear to contribute around 20\% of the total star-formation rate density at $z\sim2-3$ \citep[e.g., ][]{swinbank14} and are thus an important tracer of the star formation occurring in massive galaxies at this epoch.

The characterization of the star-forming ISM at these high redshifts is typically achieved through observations of the rotational transitions of carbon monoxide (CO) or through deep imaging of the Rayleigh-Jeans (RJ) tail of the dust continuum emission \citep[for reviews see,][]{carilli13, casey14}. However, calculations based on these ISM tracers involve a number of assumptions about the inferred gas properties.
Although CO is the most strongly emitting molecule, it is only the second most abundant molecule in the galaxy ISM after molecular Hydrogen, H$_2$, and a conversion factor ($\alpha_{\rm CO}$) from the ground-state CO($J=1-0$) luminosity to H$_2$ \citep[e.g., ][]{bolatto13} is thus required to compute the total molecular content. 
As a result there have been numerous observational \citep[e.g., ][]{tacconi08, danielson11, genzel12, hodge12, bolatto13, bothwell13, accurso17} and theoretical \citep[e.g., ][]{narayanan11, narayanan12, lagos12} efforts to constrain $\alpha_{\rm CO}$ in different galaxy populations, which represents a significant uncertainty in total gas mass estimations. 

Already in the local universe, the range of $\alpha_{\rm CO}$ values is observed to span a factor of $\sim$5, and it has been shown to be a function of several galaxy properties such as gas density, temperature, and metallicity \citep{bolatto13}, which likely evolve with redshift. 

Another approach to estimate galaxy gas masses is to use the dust continuum observations as a proxy for the ISM content at low and intermediate redshifts \citep[e.g.][]{hildebrand83, leroy11, magdis11, scoville14}.
Recently, \citet{scoville16} combined molecular gas masses inferred from existing CO detections and dust continuum measurements in an attempt to calibrate an empirical scaling factor for using global measurements of the Rayleigh-Jeans dust continuum (probed in the sub-mm regime) to estimate the total ISM masses.
Although they find that this calibration holds for measurements over three orders of magnitude in infrared luminosity and for different populations including SMGs, this is based on significant assumptions about the properties of the dust spectral energy distribution (SED) in addition to the uncertainties on the $\alpha_{\rm CO}$ parameter assumed for the calibration.
To test the validity of these assumptions, observational constraints on the physics of the high-redshift ISM are required.

Spatially unresolved high-redshift surveys of CO and dust continuum have begun to make progress on these tracers and their implications for the SF picture from a global perspective.
Over 200 detections of CO line emission at high redshift have been reported \citep[$z$ > 2;][]{carilli13} from both individual sources 
\citep[e.g., ][]{genzel12, hodge12, strandet17} and larger
statistical surveys \citep{ivison11, bothwell13, sharon16, walter16}.
In addition, multiple sub-mm continuum surveys have been conducted to characterize the dust emission and population properties of SMGs \citep{hodge13a, karim13, simpson17}.
Statistical studies \citep[e.g., ][]{bothwell13, sharon16} have shed light on the average level of excitation for SMGs, which are observed to have large scatter at high excitation levels $J\geq4$ but behave relatively homogeneously at $J\leq3$ with excitation ratios of r$_{3-1}=0.78 \pm 0.27 $ \citep{sharon16}.
High-excitation CO transitions ($J\geq 4$) have thus been observed to underestimate the gas masses, since they are biased to trace the most compact molecular emission only.
Finally, using ISM mass estimates based on dust-continuum observations and assuming a constant dust/gas ratio \citep[e.g., ][]{leroy11}, \citet{genzel15},  \citet{scoville17} and  \citet{tacconi17} were able to investigate statistically the evolution of the unresolved star-forming ISM for SF galaxies across cosmic time.

At high resolution, significant progress with the characterization of the high-redshift ISM has been achieved through the observation of gravitationally lensed sources, as part of surveys conducted by the South Pole Telescope \citep[\textit{SPT}; e.g.,][]{greve12, vieira13, spilker16} and \textit{Herschel} \cite[e.g.][]{harris12, wardlow13}. 
Lensing observations, however, are sensitive to the lensing models used to reconstruct the intrinsic morphology of the sources, and caution must be exercised for possible differential magnification biases especially when more than one ISM tracer are observed.
Only a handful of studies have characterized the ISM on sub-galactic scales in unlensed high-redshift galaxies \citep[e.g.,][]{tacconi08, engel10, hodge12, hodge13b, aravena14, decarli16, chen17}, and even fewer have studied the dust-continuum emission and gas observations on comparable scales. 
High-resolution imaging is crucial for characterizing the ISM in galaxies, since apart from providing a better morphological description of the gas or dust continuum emission, this information is key for estimating fundamental properties 
such as gas surface density. 
Moreover, through dynamical modeling of the velocity fields, one can derive dynamical mass estimates \citep{deblok14}, which reflect the total mass of baryonic and non baryonic matter contained within the region traced by the observed line emission, and thus constraints the sum of stellar, gas, and dark matter masses.
When complemented with stellar mass and dark matter fraction assumptions, dynamical mass estimates 
can constrain the total mass of the gas reservoirs \citep{tacconi08, hodge12}, and hence the $\alpha_{\rm CO}$ parameter.  

However, estimating the mass of the other components is also complex,
for example the stellar mass of a galaxy is usually inferred via SED fitting, which suffers strong degeneracies between the star formation history, dust reddening, luminosity-weighted age of the stellar populations and mass-to-light ratio, especially for highly star-forming galaxies such as SMGs \citep[e.g., ][]{ hainline11, simpson14, simpson17}.
The dark matter fraction represents a large source of uncertainty, as no independent measurement of its mass is possible.
While recent spectroscopic surveys have claimed dark matter fractions around 10--30\%, \citep{price16, wuyts16, genzel17} revealing that star-forming galaxies at $z>2$ appear to be heavily baryon dominated, these calculations involve making similar uncertain assumptions about the gas fractions. 
Shedding light on the impact of these uncertainties on the $\alpha_{\rm CO}$ and gas mass estimations is thus imperative for an improved characterization of the high-redshift ISM.

\begin{table*}
	\caption{Description of the ALMA Cycle 2 Band 3 observations and native beam properties }
		\centering

	\begin{tabular}{lcccc} 
		\hline
		&ALESS49.1 & ALESS57.1 &ALESS67.1 &ALESS122.1 \\
		\hline
		
	R.A. (J2000)&	03:31:24.72 & 03:31:51.92 & 03:32:43.20 &03:31:39.54  \\
	Dec. (J2000)&	$-27:50:47.1$ & $-27:53:27.1$ & $-27:55:14.3$ &$-27:41:19.7$  \\
	$z$ (opt)  & 2.95 & 2.94  & 2.12 & 2.02 \\
	$z$ (CO(3-2))  & $2.943 \pm 0.001 $  & $2.943 \pm 0.002$ & $2.121 \pm 0.004$ & $2.024 \pm 0.001$ \\

	    \hline
	    \multicolumn{5}{c}{Natural weighted imaging } \\
	    \hline
	Cleaning mask radius [$''$]\footnote{\label{size}Chosen as explained in Figure \ref{fig:COlines}.}.  & 1.8 & 0.8 & 2.0 & 1.0\\
    Synthesized beam FWHM [$''$]   & $0.69\times 0.63$ &$0.67\times 0.60$ & $0.56\times 0.48$& $0.45 \times 0.35$ \\
	Continuum RMS [$\mu$Jy beam$^{-1}$]  &   17.6  & 19.5 &  18.2 & 16.3\\
	Channel widths in final cubes [MHz]  & 15.6\footnote{\label{footnote:A49}Higher spectroscopic resolution was achieved in ALESS 49.1 as compared to the other sources, since the final cube is concatenated from two different measurement sets. See Section \ref{subsec:ALMAdata} for details.}&  60.5 &81.5 & 77.3\\ 
	Channel widths in final cubes [km s$^{-1}$] &54$^{\ref{footnote:A49}}$ &200 &213 &196  \\ 
    Channel RMS [$\mu$Jy beam$^{-1}$] & 182$^{\ref{footnote:A49}}$ & 148 & 113 & 131\\ 
    	\hline

	\end{tabular}
	\label{tab:native}
\end{table*}

The collecting area and sensitivity of the Atacama Large Millimeter Array (ALMA) is transforming our view of the star-forming ISM in distant galaxies.
While high-resolution studies of the dust continuum emission in SMGs with ALMA have shown this material appear to be mostly distributed in compact regions \citep{simpson15, ikarashi15, hodge16}, the extended sizes revealed by the few resolved CO detections of luminous sources \citep{riechers10, hodge12, emonts16, chen17, dannerbauer17, ginolfi17} challenge general assumptions of co-spatial interwoven dust and gas generally assumed by models.
Spatially resolved observations of CO and dust continuum emission for homogeneously selected samples and modelling their interplay, e.g. through radiative transfer approximations, may help characterize the distributions and physics of dust and gas in the high-redshift ISM. 

Here, we present high-resolution imaging of the CO emission in four SMGs from the ALESS survey \citep{hodge13a, karim13} at sub-arcsecond resolution. 
We address the following questions, which remain open in ISM studies at high redshift: 
How is the molecular gas distributed in relation to the dusty ISM and the stellar populations?
What implications do these distributions have for the assumptions made for the dust spectral energy distributions and dust-to-gas ratios at high redshift?
How reliable are gas mass estimations?
What uncertainties do stellar mass estimates and dark matter fraction assumptions introduce into the total gas mass estimation?
The paper is structured as follows: in Section \ref{sec:data} we describe the ALMA data reduction and the imaging of the CO(3-2) maps. In Section \ref{sec:molgasdyn} we present the analysis of the kinematic properties of the CO(3-2) emission in our sources and present the implications of these to total gas estimates. In Section \ref{sec:distributions} we present an statistical analysis of the distributions of gas, dust continuum and stellar emission and discuss the physical implications of these findings.
Throughout the paper we adopt a $\Lambda$-CDM cosmology, consistent with the values given by the \citet{planck14}, with $\Omega_{\Lambda}$ = 0.69, $\Omega_{m}$ = 0.31 and $H_{0}$ = 67.7 $\rm km \, s^{-1} Mpc^{-1}$.


\section{Observations and Imaging}
\label{sec:data}

\subsection{Target selection ALMA observations} \label{subsec:ALMAdata}

We present ALMA Cycle 2 observations of the CO emission from four SMGs, ALESS49.1,  ALESS57.1,  ALESS67.1 and ALESS122.1.
These sources were selected from the ALESS survey \citep{hodge13a, karim13}: an ALMA Cycle 0 follow-up program of 126 sources detected in the single-dish LABOCA Extended \textit{Chandra} Deep Field South (ECDFS) Submm Survey \citep[LESS,][]{weiss09}. 
These four sources were selected as they are spectroscopically confirmed at redshifts $2.1 <z<2.9$ as the result of an extensive spectroscopic follow-up campaign  \citep{danielson17}. 
Three of the four sources in our sample were detected (ALESS57.1 and ALESS67.1) or marginally detected (ALESS122.1, through stacking) in the X-rays \citep{wang13}, using 4 Ms $Chandra$ observations of the CDFS region \citep{X11} and 250 ks observations in the ECDFS \citep{L05}. 
ALESS57.1 and ALESS122.1 are reported as AGN candidates \citep{wang13, danielson17}, while the origin of the X-ray emission in ALESS67.1 is most probably related to star formation.

ALMA observations were taken in Cycle 2 as part of the project 2013.1.00470.S (PI: Hodge), with a total integration time of $\sim$2.5 hours in ALMA Band 3, covering the spectral range  expected for the line emission of the CO(3-2) transition at these redshifts.
The sources were observed on 2015 September 4, 6 and 20, using the 12-meter array and under good phase stability/weather conditions, with PWV at zenith ranging between 1.56--3.03 mm. 
The antenna configuration consisted of 33, 36 and 35 antennas, respectively, achieving synthesized beam sizes that range between 0.34 -- 0.67$''$ major axis FWHM with largest angular scales (LAS) between 1.9--2.9$''$. 
The observations were calibrated based on Jupiter as the flux-calibrator,  J0334-4008 as the band-pass calibrator, and J0334-401 as the phase calibrator.

New ALMA observations of ALESS49.1 (Wardlow et al. 2018) were taken during Cycle 4 on 2016 November 12, 16 and 20 as part of the project 2016.1.00754.S (PI: Wardlow).
These observations were carried out using a total integration time of $\sim 2700$ s and using the longest baselines of $\sim$ 650 m.
With an angular resolution of 1.1$''$, these additional data increase the signal-to-noise ratio (SNR) of our high-resolution data.
We concatenated both Cycle 2 and Cycle 4 datasets achieving a combination of high-resolution imaging, high SNR and at the same time reducing concerns whether we might have resolved out any significant flux from ALESS49.1, thanks to the short baselines covered.
An analysis of the environment of ALESS49.1 is presented in Wardlow et al. (2018).

The ALMA data were calibrated following the ALMA pipeline and using the Common Astronomy Software Application package \citep[CASA, ][]{CASA}.
Manual flagging of a few corrupted time windows for ALESS57.1 was required after an inspection of the calibration output.
The imaging process was carried out using CASA tasks (version 4.7.0).
The uv-data were Fourier transformed and deconvolved from the point spread function using the CASA \textsc{clean} algorithm. 
After resampling of the visibilities at different spectral resolutions to optimize the SNR, we produced the final data cubes averaging the visibilities to channel widths of 16, 61, 82 and 77 MHz for ALESS49.1, 57.1, 67.1, 122.1, respectively.
At the native resolution, the rms values achieved for the final data cubes range from 0.11-0.18 mJy beam$^{-1}$ (see Table~\ref{tab:native}).

\begin{figure*}
\centering

	 \includegraphics[width=0.87\textwidth]{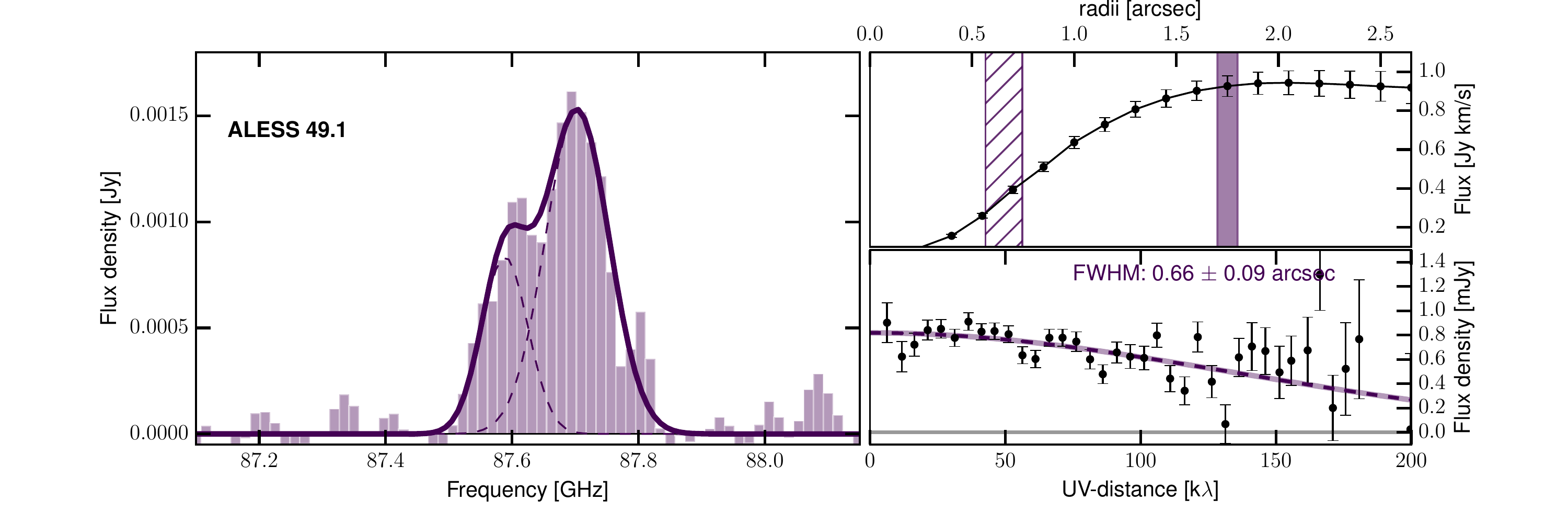}

    \includegraphics[width=0.87\textwidth]{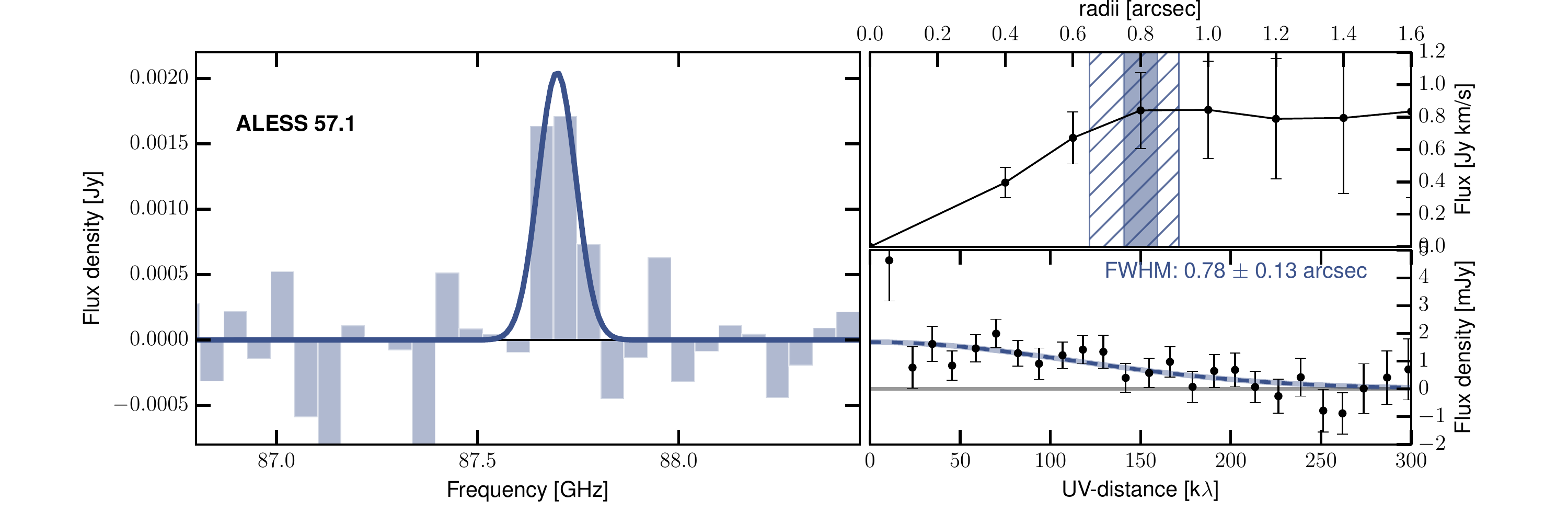}

    \includegraphics[width=0.87\textwidth]{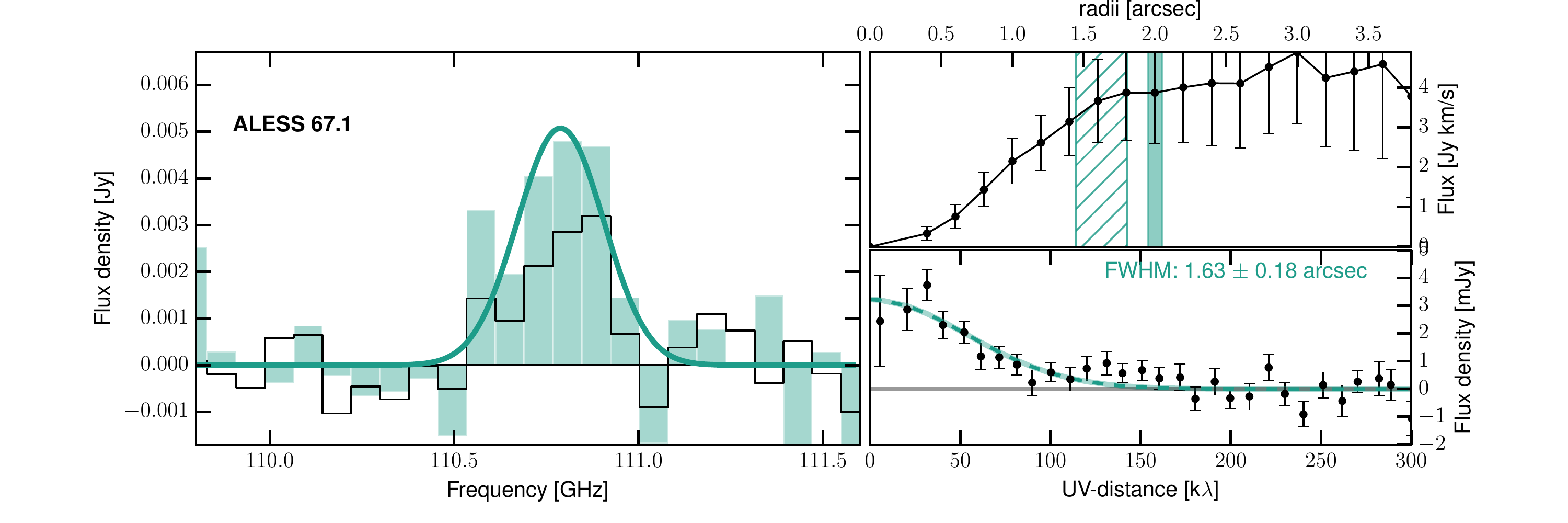}

    \includegraphics[width=0.87\textwidth]{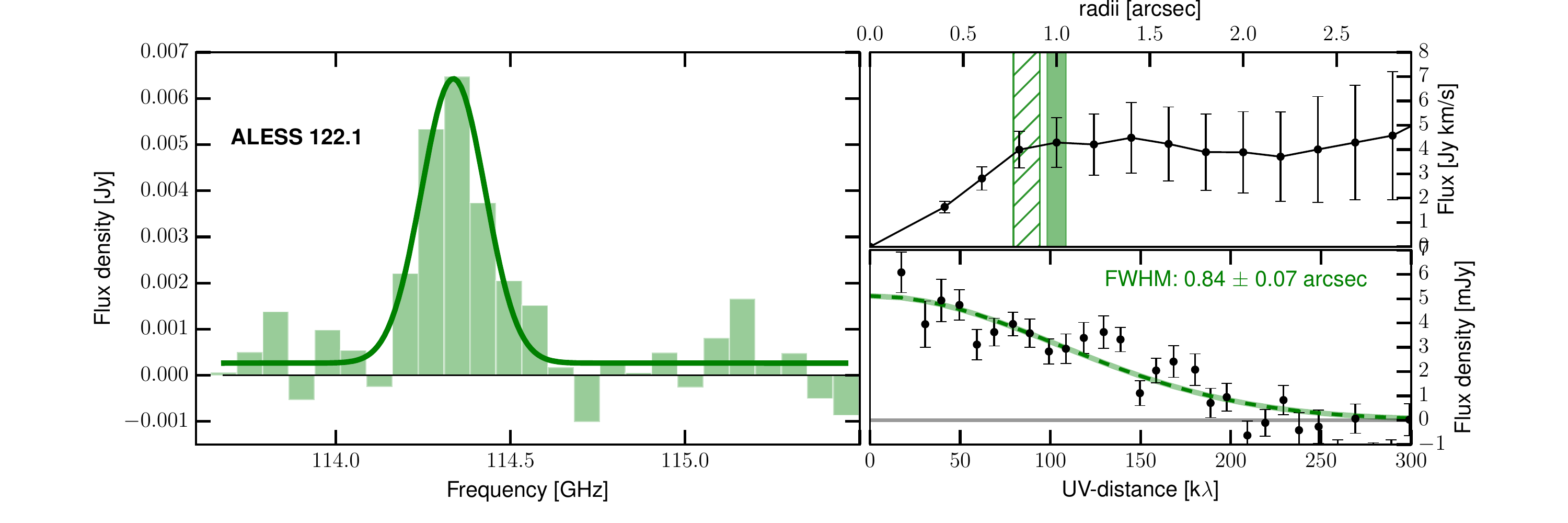}

    \caption{\textbf{Analysis of the CO(3-2) data:} The\textit{ left panels } show the CO(3-2) spectra of our sources. The  spectral resolution is $\sim$54, 200, 213, and 116 km~s$^{-1}$, for ALESS49.1, ALESS57.1, ALESS67.1 and ALESS122.1,  respectively. Gaussian fits to the spectra are shown by the solid lines. For ALESS49.1 we fit a two-component Gaussian profile and show each component as dotted lines. For ALESS67.1, we additionally show the CO spectrum extracted exclusively from within a mask of $\sim 1"$ around the centroid of the main component as a black line (see Section \ref{subsec:COline} for more details). The \textit{right panels} show two methods used to calculate the cleaning masks and intrinsic source sizes, respectively. The first method (\textit{top right}) is a curve-of-growth analysis, where the shaded area shows the radius at convergence. This method was used to determine the size of the area used for masking the cleaning process. The second method (\textit{bottom right}) is the analysis of the visibilities (uv)-profiles, which reliably estimate intrinsic source sizes. A single Gaussian fit to the uv-data is shown by the line.
    The half-light radius $r_{1/2}$ from the uv-fitting method is a more robust estimate of the true size of the sources, since the curve-of-growth analysis is prone to be affected by correlated noise. We draw hashed areas to show the FWHM from the Gaussian fit (i.e., $2\times r_{1/2}$) on the top-right panel and  see that in most cases, it roughly corresponds to the peak of the curve of growth.}
    
    \label{fig:COlines}
\end{figure*}

Due to the high spatial resolution of the data, it is not trivial to estimate the masks on which to apply the \textsc{clean} task.
We adopted an iterative cleaning technique \citep[e.g., ][]{chen17}, in order to optimize the mask size estimation to include possible extended low surface brightness emission. 
Iterative cleaning consists of drawing concentric circular mask regions at increasing radii and applying the \textsc{clean} task and line flux extraction within them.
Plotting the resulting line fluxes against the corresponding circular region radii, a curve of growth is produced (see upper right panels of Fig. \ref{fig:COlines}).
The expected behaviour of the measured flux density in a curve of growth is to continuously increase as a function of radius, reaching a point of convergence at the maximum extent of the source.
We used the masks inferred from the iterative cleaning method to extract the final line cubes.
We explored the data for emission at different spatial scales and surface brightness, first at the native resolution using natural weighting, then using a Briggs weighting with a robustness parameter 0.5, to image the CO emission at lower SNR but slightly higher spatial
resolution and finally, by tapering the visibilities to lower resolutions (1.5-2 $\times$ native beam size) to recover extended emission.

\begin{figure*}
\centering

	\includegraphics[trim={0.25cm 0 0cm 0cm},clip,width=0.35\textwidth]{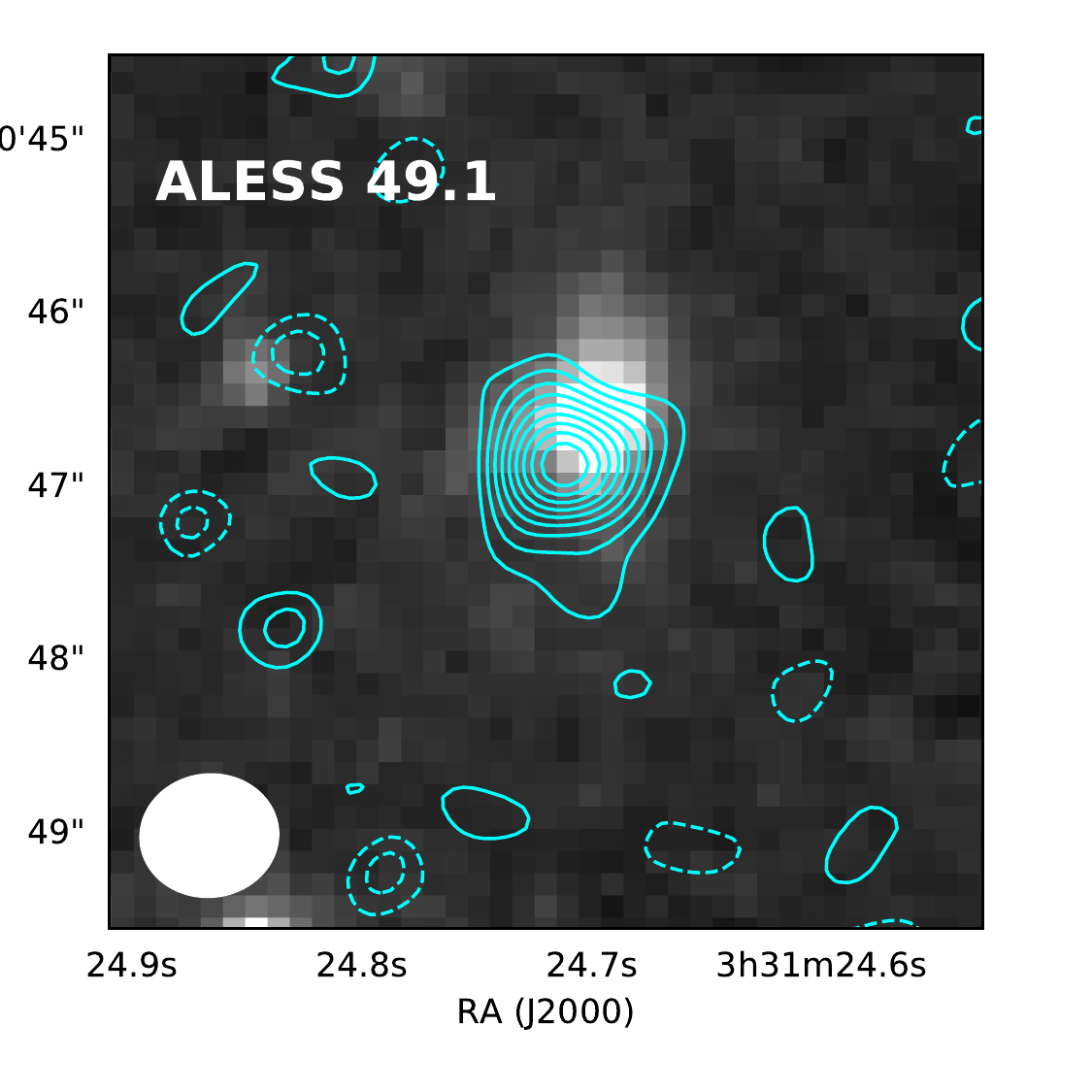}
	 \includegraphics[trim={0.25cm 0 0cm 0cm},clip,width=0.35\textwidth]{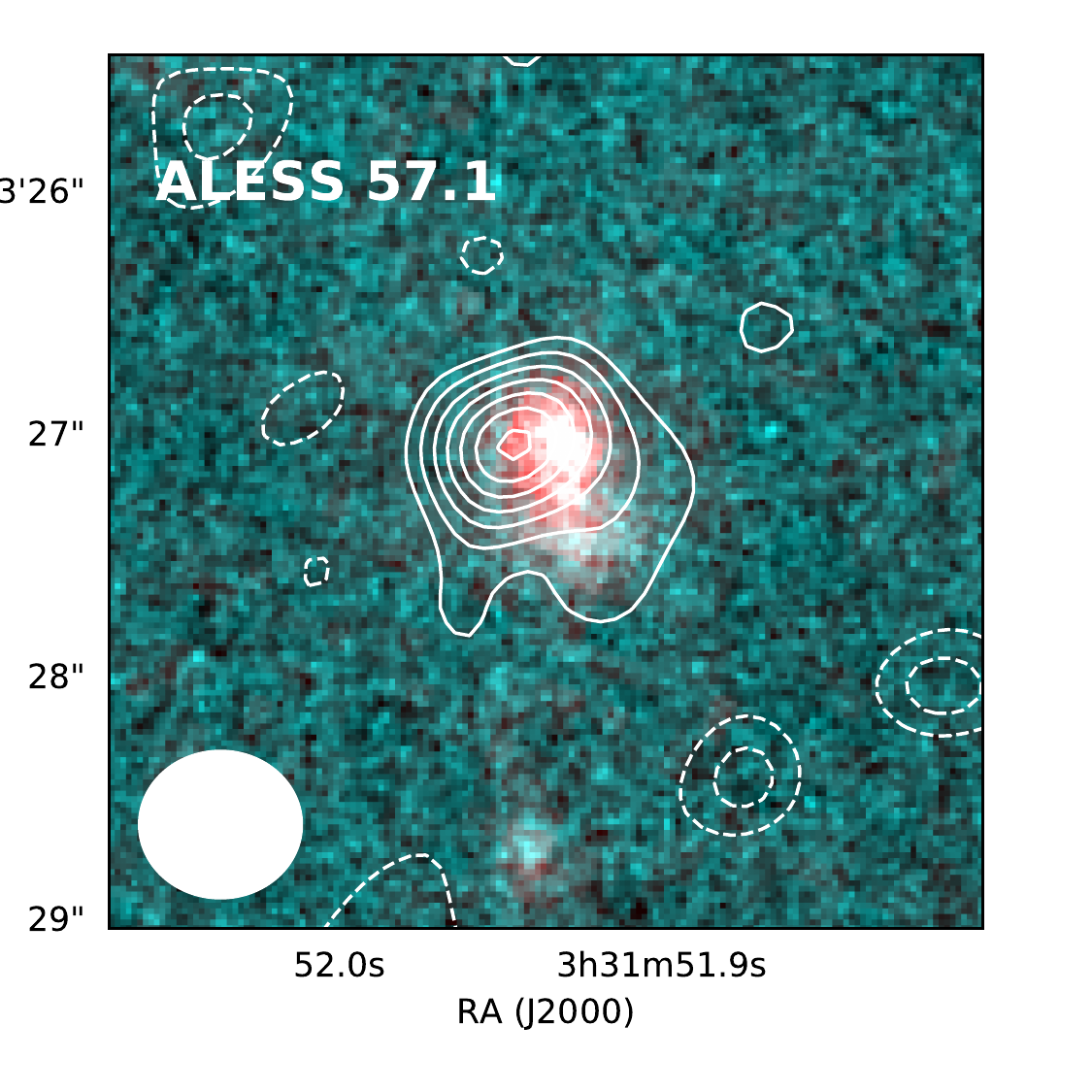}\\
  \includegraphics[trim={0.25cm 0 0cm 0cm},clip, width=0.35\textwidth]{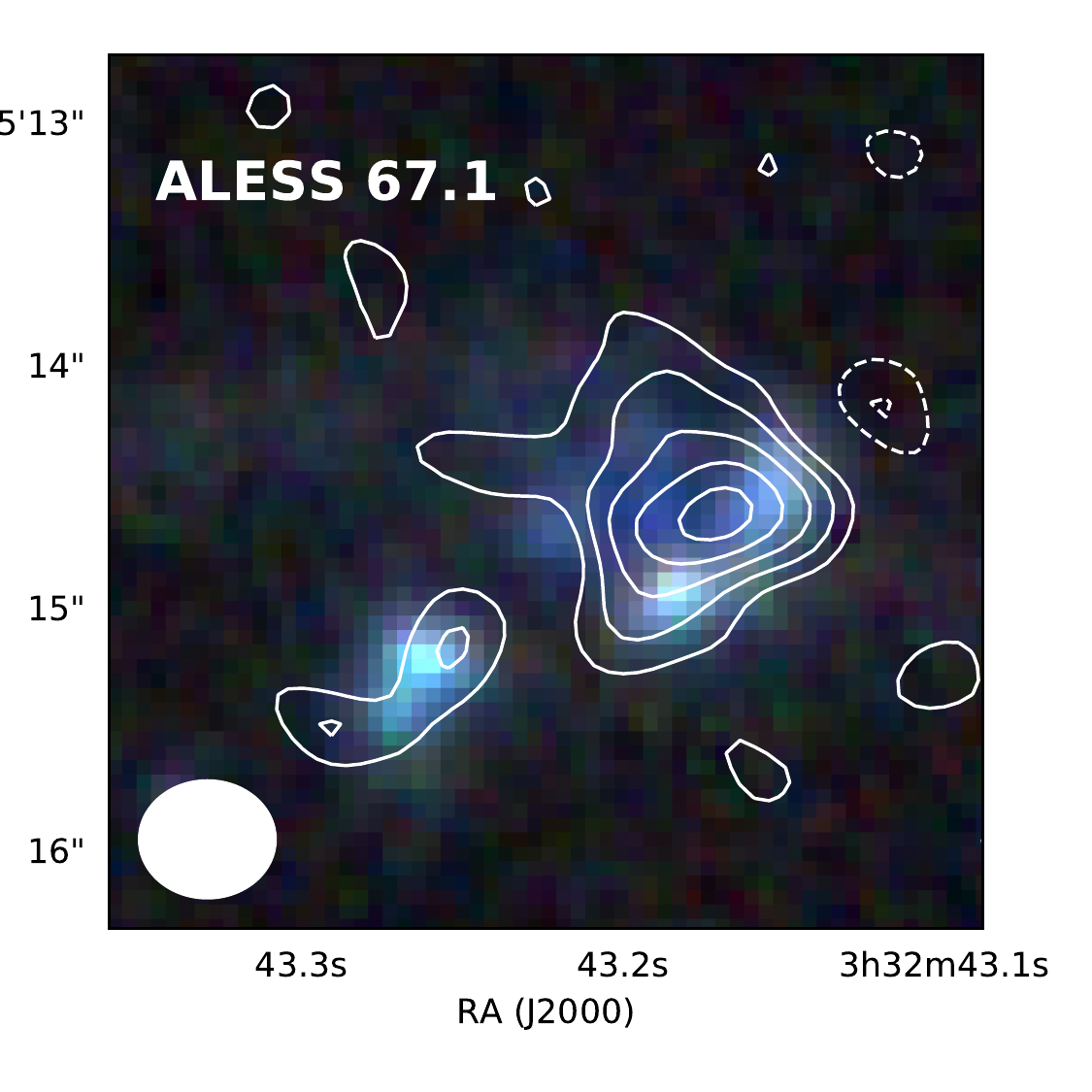}
\includegraphics[trim={0.25cm 0 0cm 0cm},clip, width=0.35\textwidth]{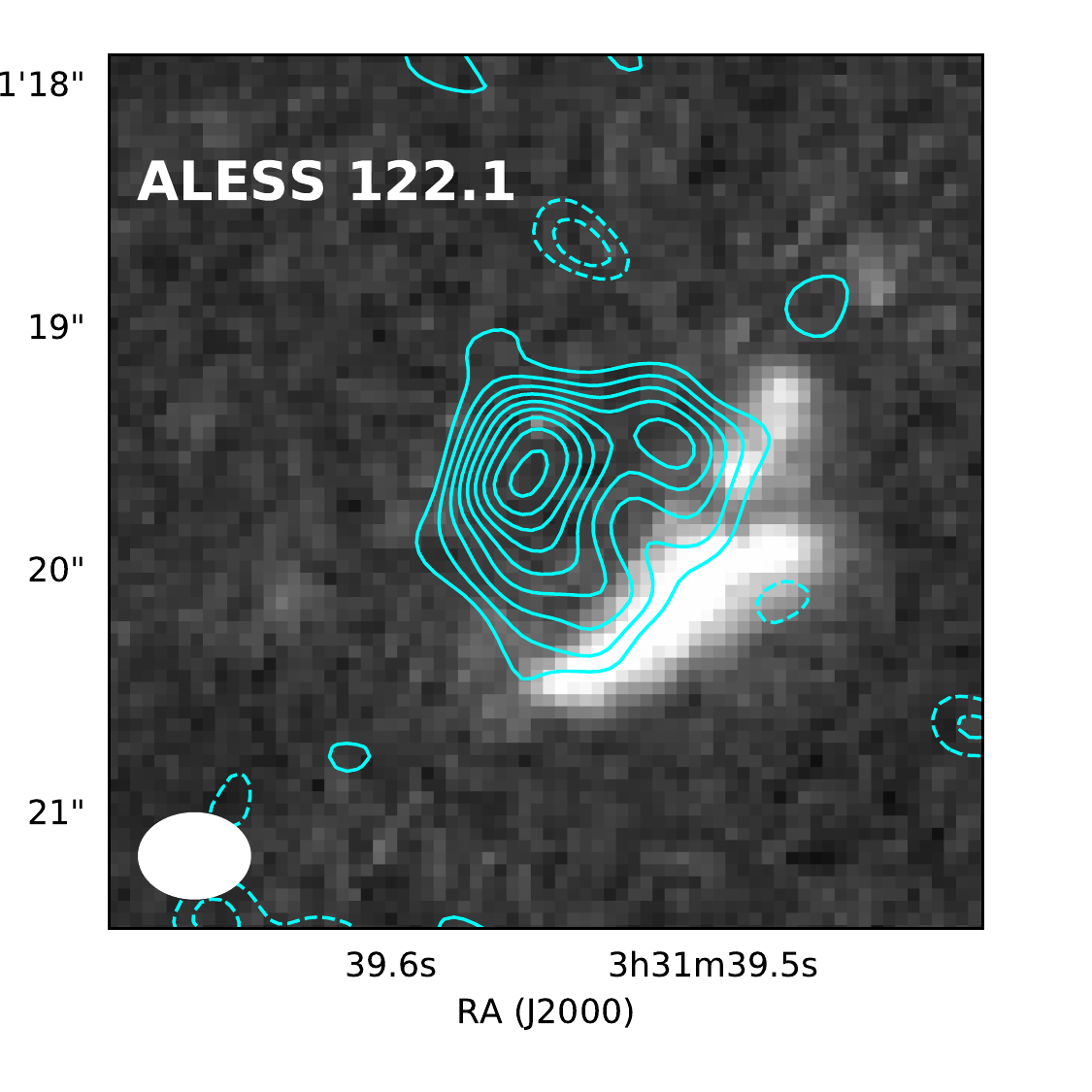}
    
    \caption{CO(3-2) velocity averaged contours overlaid on optical/near-IR cutouts from the available \textit{HST}-WFC3 and/or ACS imaging. Resolutions of 0.4--0.6$''$ are achieved through natural weighting. The contours are obtained by averaging over the velocity range corresponding to the FWHM of the Gaussian fit to the CO spectrum.  The contours in all maps show 2-10 $\sigma_{\rm rms}$ regions, where  the $\sigma_{\rm rms}$ for the velocity averaged images is:  ($0.10$ mJy beam$^{-1}$) for ALESS49.1,  ($0.15$ mJy beam$^{-1}$) for ALESS57.1,  ($0.11$ mJy beam$^{-1}$) for ALESS67.1 and  ($0.13$ mJy beam$^{-1}$) for ALESS122.1. The beam sizes are shown in the lower-left corners of the images. The background map  for ALESS49.1 represents the $H_{160}$ imaging, the two-color map for ALESS57.1 uses the $H_{160}$ and $ I_{814}$ fluxes, the background r-g-b map for ALESS67.1 represents $H_{160}, J_{125}, Y_{105}$ and the background map for ALESS122.1 shows the $ I_{814}$ imaging. The \textit{HST} imaging has been corrected for astrometric offsets using  \textit{GAIA} data. We see that our observations strongly detect and resolve the CO emission of these sources on scales of ~3--5 kpc, finding different morphologies in both CO and stellar emission.}
    \label{fig:velaver}
\end{figure*}

\subsection{CO(3-2) line detections and continuum} \label{subsec:COline}

We detect significant CO(3-2) line emission for the four sources in our sample even from the dirty data cubes, i.e. in the Fourier transformed visibilities prior to the beam deconvolution (cleaning process).
The CO detections are coincident with the expected ALMA Cycle 0 positions, and at frequencies which confirm the previously reported spectroscopic redshifts (Table \ref{tab:native}).
We identify strong line detections for all the sources in the final line cubes constructed by applying the circular cleaning masks described in Section \ref{subsec:ALMAdata}.
Fig. \ref{fig:COlines} show the data as well as Gaussian profile fits to the extracted spectra.
The estimated CO(3-2) line parameters are summarized in Table \ref{tab:COprops}. 
We next briefly describe some important features of the line emission of each source. 

The CO spectrum of ALESS49.1 is clearly best described by a double-peaked profile and was thus fitted using an asymmetrical two-Gaussian model. 
The red and blue component line widths are $280\pm40$\kms and $390\pm30$ \kms (FWHM), respectively, and are separated by $300 \pm 20$ \kms. 

The CO spectrum of ALESS57.1 is marginally spectroscopically resolved. 
Nevertheless, the line width is appropriately approximated by a Gaussian fit and the resulting values are summarized in Table \ref{tab:COprops}.

The CO line of ALESS67.1 shows a very large line width (FWHM$\sim 800$~km~s$^{-1}$) and appears asymmetric although the asymmetry is not statistically significant, due to the low SNR in this source.
We postulate that the CO line of ALESS67.1 is produced by two sources of emission.
This is supported by the velocity averaged image (Fig. \ref{fig:velaver}), where two spatial components are recognizable independently in both CO (ALMA) and optical/near-infrared (\textit{HST}) imaging: one main right component of around 1 arsec radius and a second extended emission in the east. 
We thus give preference to the description of this CO line as originating in two sources (possibly an early stage of a merger event) and focus on the emission extracted from the main component (1 arsec radius), which is shown as a black line overlaid on the total line emission in Figure~\ref{fig:COlines}.

The CO spectrum of ALESS122.1 is the brightest in our sample, and its line profile is well described by a single-Gaussian fit. Table \ref{tab:COprops} lists the CO(3-2) line properties and best-fit parameter values.
A detailed description of the velocity field sampled by these CO lines will be provided in Section \ref{subsubsec:channelmaps}.

Continuum images were made for all the sources after excluding the channels contributing to the CO(3-2) line emission.
Using natural weighting to achieve the highest sensitivity\footnote{\label{footnoteLowRes}The rms for ALESS49.1 is achieved using the concatenated dataset which has a lower resolution ($\theta_{\rm beam}\sim$0.7$''$) than the Cycle 2 only ($\sim$0.6$''$), as discussed in Section \ref{sec:data}. The 88 GHz continuum emission measured from the Cycle~4 low resolution data alone ($\theta_{\rm beam}\sim$1.5$''$) has a $\sim 9 \sigma$ detection},
we obtained images with $\sigma_{\rm rms}\sim 6$ 
$, 20, 18, 12$ $\mu$Jy beam$^{-1}$, as summarized in Table \ref{tab:native}. 
The continuum emission is detected in only two of the four sources in our data set: ALESS49.1 and ALESS122.1. 
In both cases, the naturally weighted dust continuum emission at 3mm (rest-frame $\sim$850 $\mu$m) is detected at 4-$\sigma$ significance at sub-arsecond resolution (Table \ref{tab:COprops}).

To test for consistency with the Cycle 0 continuum observations at $870 \mu$m \citep{hodge13a}, we use an extrapolation following $S_{\nu} \propto \nu^{2+\beta}$, where $S_{\nu}$ is the measured flux density and $\beta$ is the dust emissivity spectral index.
Adopting $\beta$ = 1.5
as a reasonable assumption for dusty galaxies \citep{swinbank14, casey14}, 
the values extrapolated from $870 \mu$m  are consistent with our measured $S_{\rm 3mm}$ peak flux densities and upper limits.

\begin{table*}

	\centering
	\caption{CO(3-2) line properties and estimated quantities.}
	\begin{tabular}{lcccc}
		\hline
		
		&ALESS49.1 & ALESS57.1 & ALESS67.1 & ALESS122.1 \\
		\hline
		
	CO(3-2) $r_{1/2}$ [kpc]	& $2.6 \pm 0.4$& $3.1 \pm 0.5 $ & $6.9 \pm 0.8$& $3.6 \pm 0.3$ \\
   FWHM$_{\rm CO(3-2)}$[km s$^{-1}$] &$610\pm30$ &$360\pm90$ &$720\pm160$ ($500\pm110$)\footnote{\label{footnote3} Values restricted to the main component of ALESS67.1 shown in Fig. \ref{fig:COlines}. These values were used for the calculation of the dynamical masses using the disk approximation in Section \ref{subsec:dynamicalmasses}.} &$600\pm80 $\\
   Dust continuum $S_{\rm 3\,mm} [\mu$Jy]	& $34 \pm 6$& $54\pm18$ & $\leq 54$\footnote{\label{footnote4}3$\sigma$ upper limits.} & $\leq 60^{\ref{footnote4}}$ \\

	Inclination	$i$[degrees]\footnote{Computed from axial ratios estimated with the CASA task \textsc{imfit}} & $ 80\pm30$&  $40\pm50 $ & $ 50\pm20$& $40\pm20 $ \\
	Velocity integrated flux density $I_{\rm CO(3-2)}$ [Jy km s$^{-1}$]	&$0.89 \pm 0.07$& $0.8\pm0.2$ & $3.9\pm1.2$ ($1.8\pm0.5)^{\ref{footnote3}}$& $4.2 \pm 0.8$\\

	Velocity integrated flux density $I_{\rm CO(1-0)}$ [Jy km s$^{-1}$]	&--&--&
	$0.44\pm0.08$ \footnote{\label{footnote2}Measured by \citet{huynh17}.}& 
	$0.64\pm 0.07 ^{\ref{footnote2}}$\\
	CO(3-2) line luminosity $L'_{\rm CO(3-2)}$ [$10^{11}$ K km s$^{-1}$ pc$^{2}$]
&$ 0.39 \pm 0.03$
&$0.4 \pm 0.1$&
$1.0 \pm 0.3$ $(0.5 \pm 0.1)^{\ref{footnote3}}$&
$1.0 \pm 0.2$\\

	CO(1-0) line luminosity $L'_{\rm CO(1-0)}$ [$10^{11}$ K~km s$^{-1}$~pc$^{2}$]
	&$0.5\pm 0.2 $\footnote{\label{footnote}Estimated using $r_{3/1}=0.78 \pm 0.27$ \citep{sharon16}.} & 
	$0.5 \pm 0.2^{\ref{footnote}}$ &
	$1.0\pm0.2^{\ref{footnote2}} $
	&$1.3 \pm 0.2^{\ref{footnote2}} $\\
  	Molecular gas mass [$10^{11} \rm M_{\odot} (\alpha_{\rm CO}=1.)$ ]
  	& $0.5 \pm 0.2$& $0.5 \pm 0.2$&$1.0\pm0.2$&$1.3 \pm 0.2 $\\
   $M_{\rm dyn}(r \leq 2 r_{1/2})$ [$10^{11}$ M$_{\odot}$] & $1.1 \pm 0.2$ & $1.1 \pm 0.5$ & $3.6 \pm 1.6$& $5.3 \pm 1.6$\\ 
	\hline
  	Stellar mass$^{\ref{dacunha}}$  [$10^{11} \rm M_{\odot}$ ]
  	&$ 0.4 \pm 0.1 $&$0.8\pm 0.1$&$2.4 \pm 2.1 $&$0.8 \pm 0.5 $\\
    Infrared luminosities $L_{\rm IR(3-2000~\mu m)}[\times 10^{12}L_{\odot}]$\footnote{\label{dacunha}As presented by \citet{dacunha15}}	& $6.8 \pm 0.6$& $2.3\pm2.2$ & $5.0\pm1.5$ & $8.3\pm2.5$ \\
 	SFR$^{\ref{dacunha}}$  [$\rm M_{\odot}~yr^{-1}$ ]
  	&$ 700 \pm 100 $&$200\pm 200$&$400\pm 100 $&$700 \pm 200 $\\
	\hline

	\end{tabular}
\label{tab:COprops}
\end{table*}

\subsection{Source size estimation}
\label{subsec:sourcesizes}

High-resolution imaging reveals the detailed structure of the gas and dust continuum emission, which is key for placing constraints on the dynamical states of the sources.
However, high-resolution interferometry is less sensitive to extended, low surface brightness emission, and thus care must be exercised in the estimation of the source sizes \citep[e.g.][]{emonts16, dannerbauer17, ginolfi17}.
Although the iterative cleaning method presented in Section \ref{subsec:ALMAdata} can provide a sense of the \textit{total} source extent, it is prone to correlated noise effects and consequently may yield uncertain results.
Moreover, iterative cleaning can only provide an (intrinsic) source size estimate when the extent of the source is greater than the synthesized beam. 
To determine the sizes independently of these possible beam-convolution effects intrinsic to the imaging process, we estimate the sizes directly from the uv-data (lower right panels of Figure \ref{fig:COlines}).

We extract the uv-data corresponding to the frequencies within the FWHM of the line from the continuum-subtracted cube, centered at the CO(3-2) observed frequency.
For each source, we then average the visibilities at different uv-distances and plot the resulting amplitudes against them, as shown in Fig. \ref{fig:COlines}.
We fit a Gaussian profile to the data and measure the radius within which half of the light of the galaxy is contained.
We report the estimated half-light radii $r_{1/2}$ in Table \ref{tab:COprops} and adopt these values for our analysis. 
The half-light radii of our sources range between 0.4--0.8$''$, which correspond to 2.6--6.9 kpc at the respective redshifts, implying total physical sizes (FWHM) of $\sim 10$ kpc.
In a study of the CO(3-2) emission of a similar sample of SMGs at $z\sim2-3$, \citet{tacconi08} find sizes ranging between 2--12 kpc.
More recent high-resolution molecular gas studies have reported similar extents \citep[e.g., ][]{riechers11e, ivison11, hodge12}, suggesting that low-$J$ transitions can show larger extents than high-$J$ CO emission.

Figure \ref{fig:velaver} shows the CO(3-2) emission overlaid on the stellar emission as traced by the WFC3/IR (bands $H_{160}, J_{125}, Y_{105}$) and/or ACS imaging ($ I_{814}$) from \textit{HST} \citep[e.g., ][]{chen15}.
We note that the astrometry in the \textit{HST}-images has been corrected based on the \textit{GAIA} star positions, which are aligned to the Fifth Fundamental Catalogue (FK5) \citep{gaia16}.
With the exception of ALESS122.1, the CO gas overlaps with the stellar distributions, although they have slight offsets ($0.1-0.3$", 1--3 kpc offsets).
The sizes of the gas and stellar distribution in these sources are also roughly similar. 
Specifically, the ratio of the \textit{HST} $H_{160}$ imaging \citep[][]{chen15} to CO half-light radii of these ALESS sources ranges between 1 and 1.5 (ALESS122.1 not included here as no $H_{160}$ is available). 
We will compare the extent of the different components in a statistical study for SMG populations in Section \ref{subsec:stackedprofiles}. 
ALESS122.1 displays the striking feature that the gas and stellar emission are completely offset. 
We must point out, however, that the stellar emission in ALESS122.1 is exclusively represented by the ACS $I_{814}$ band imaging (rest-frame 270 nm), in contrast to the other three sources which are covered by at least one WFC3/IR band. 
This particular case will be discussed in detail in Section \ref{subsec:offset}.

Previous studies have shown that submillimeter continuum observations of similar galaxy populations reveal very compact rest-frame far-infrared-emitting regions, with median half-light radii of only $\sim$ 0.7--1.6 kpc \citep[e.g., ][]{simpson15, ikarashi15, hodge16, oteo17}.
This is significantly more compact than the molecular gas emission in our sources.
This difference in compactness is similarly observed between the dust continuum and radio emission of SMGs. 
Median radii of radio emitting regions originating from SMGs have been reported to be around 2.1 kpc in average \citep[e.g., ][]{chapman04, biggs08, miettinen15, miettinen17, thomson18}.
Motivated by these differences, we will explore in detail the relation between the radial profiles of the molecular gas, the dust continuum emission (at rest-frame 250$\mu$m) and stellar emission in Section \ref{subsec:stackedprofiles}.

\section{Dynamical constraints to the total gas masses} \label{sec:molgasdyn}
\subsection{Molecular gas masses}

The masses of the CO(3-2) emitting gas in our sources can be estimated from the observed $^{12}$CO($J$=3-2) line luminosities $L'_{\rm CO}$.
These were calculated following \citet{solomon05} as $L'_{\rm CO}=3.25 \times 10^7 \, \rm S_{\rm CO} \Delta v \, \nu_{\rm obs}^{-2} \,D_{\rm L}^2
(1+z)^{-3} ~\mathrm{K~km~s^{-1} pc^2}$; the resulting values are listed in Table \ref{tab:COprops}. 
Through an extrapolation from these CO masses we can calculate the total molecular content of the systems (dominated by H$_2$), by assuming a CO line luminosity to H$_2$ (+He) mass conversion factor, $\alpha_{\rm CO}$, and a brightness temperature ratio of CO($J$=3-2) to CO($J$=1-0).
However, the combination of spatial and spectroscopic resolution of our data allow us to make an estimate of the total gas mass independently of the above mentioned assumptions, by estimating kinematic parameters (Section \ref{subsec:kinematics}) and using further multiwavelength information to estimate the stellar mass contribution (Section \ref{subsec:covariance}).
But to begin with we use the classical approach to estimate the molecular gas masses as a function of $\alpha_{\rm CO}$.

Using the measurements of $^{12}$CO($J$=1-0) by \citet{huynh17} for ALESS67.1 and ALESS122.1 (see Table \ref{tab:COprops}), we calculate their brightness temperature ratio $r_{3/1} = L_{\rm CO(3-2)}/L_{\rm CO(1-0)}$, which yields $r_{3/1} = 1.01 \pm 0.36$ and $r_{3/1} = 0.77 \pm 0.19$, respectively. This is consistent with previous estimates for the SMG population \citep{ivison11, bothwell13, sharon16}.
For ALESS49.1 and ALESS57.1 we estimate the $^{12}$CO(1-0) emission using the excitation ratio for SMGs derived by \citet{sharon16}, $r_{3/1} = 0.78\pm0.27$, yielding $L'_{\rm CO(1-0)}= (0.5 \pm 0.2) \times 10^{11}$ K km s$^{-1}$ pc$^{2}$, for both sources.
The derived molecular gas masses as a function of $\alpha_{\rm CO}$ (equivalent to $\alpha_{\rm CO}=1$) are reported in Table \ref{tab:COprops}.

In order to investigate possible effects of the presence of an AGN on the molecular gas in our sample, we explore possible correlations between galaxy properties inferred from the CO measurements (listed in Table \ref{tab:COprops}) such as the FWHM of the CO line, the gas-to-stellar mass fraction and star formation efficiency ($SFE = SFR/M_{\rm gas}$) with the presence or absence of an AGN (ALESS57.1 and ALESS122.1 are the AGN-host candidates in our sample, see Section \ref{subsec:ALMAdata}).
We find no clear correlation of these properties and conclude that the scales probed by our CO observations are not affected by the presence of an AGN, especially since we have observed the CO ($J=3-2$) transition, which is not expected to be a tracer of AGN excitation \citep{sharon16}.


\subsection{CO line kinematics}\label{subsec:kinematics}

The morphological and kinematic properties of our resolved CO images allow us to explore the nature of these star-forming systems, whether these are in an early stage of a merger with a chaotic structure, or whether their velocity fields are described by an ordered rotating disk.
It is important to note, however, that these scenarios are not mutually exclusive. Observing dynamics consistent with a rotating disk does not preclude the galaxy from being a merging system, as the gas is collisional and an ordered rotating disk can quickly reform after the final coalescence stage as has been shown observationally and theoretically \citep{robertson06, hopkins09, hopkins13, ueda14}.

While the few existing examples of CO detections at similar resolution suggest a mixture of mergers and disk-like motion \citep[e.g.,][]{tacconi08, engel10, hodge12},
recent high resolution continuum imaging of SMGs has shown that the dust continuum emission (at rest-frame 250 $\mu$m) is mostly described by compact disk profiles \citep[e.g.][]{simpson15, hodge16}.
Here we use our high-resolution CO observations to directly test the kinematics of these systems.


\begin{figure*}
\centering

\includegraphics[trim={3.5cm  0.cm 1cm 0cm},width=\linewidth]{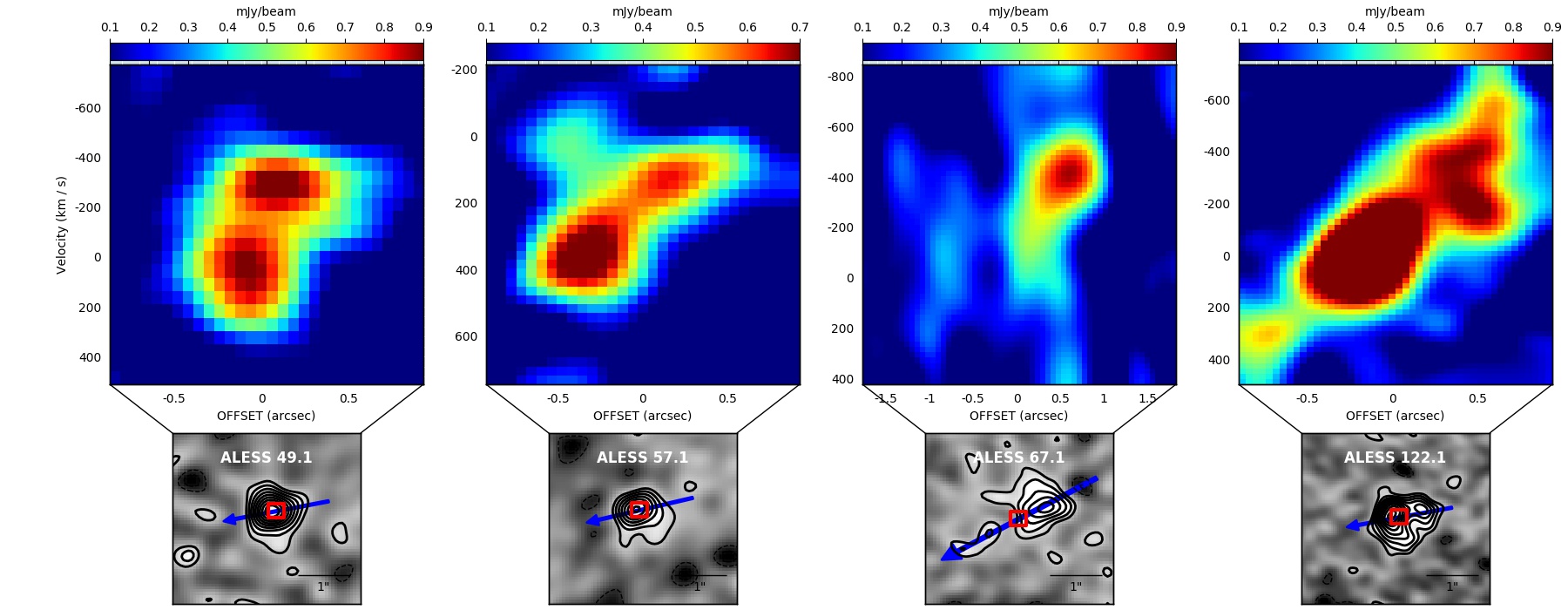}
    \caption{Position--velocity (PV) diagrams for the CO(3-2) emission in our sources. The \textit{lower panels} show the velocity-averaged CO images of our sources. The arrows represent the kinematical major axis chosen for the extraction of the PV diagrams, the arrow head shows positive velocity offset, and the squares show the origin of the PV plot. The upper panels show the corresponding resulting PV diagrams for the respective sources. Velocity gradients from the south-west to north-east direction can be recognized in at least three sources (ALESS 67.1 may also have weak evidence), suggesting that the CO emitting material in these sources is supported by rotation with velocity gradients of $\sim$300 \kms on spatial scales of $\sim 0.2$--$0.7''$.}
    \label{fig:channelmaps}
\end{figure*}

\subsubsection{Position--velocity diagrams}\label{subsubsec:channelmaps}

To estimate the kinematic properties of our sources we first investigate their position--velocity (PV) diagrams.
We apply the interactive PV Diagram Creation task from the CASA \textsc{viewer} to the data cubes of our sources with spectral resolution corresponding to channel widths of $\sim 100$\kms.  
The lower panel of Figure \ref{fig:channelmaps} show the kinematical major axis chosen for the extraction of the PV diagrams overplotted as blue arrows on the velocity-averaged CO images. 
The upper panels show the resulting PV diagrams for the respective sources.
We find velocity gradients in the PV diagrams of three of our four sources (with ALESS67.1 also showing some weak signs), suggesting that the bulk of emission in these sources is dominated by rotation.

The velocity gradients of ALESS49.1, ALESS57.1 show double peaks at each side of the galactic center, although the detailed velocity structure is ambiguous given the modest SNR.
In ALESS49.1, the double-peaked spatial structure gives rise to the double-peaked line profile observed in Figure \ref{fig:COlines}.
Since both sources have the CO centered on a single optical nucleus (as seen from their high-resolution \textit{HST} imaging), the velocity structure suggests the presence of a disk structure or a disk-shaped merger remnant, rather than an early stage merger.

The CO emission in  ALESS67.1  and its velocity structure appears more chaotic, which is probably accentuated by the combination of extended emission and the lower SNR achieved in this source. 
A detailed analysis of the velocity structure of this individual source has been conducted in a complementary study by \citet{chen17}.
They presented \textit{SINFONI}~$H\alpha$ observations of ALESS67.1, revealing that the extended CO emission (including the second component to the south-east) follows the bulk rotational motion of the rest-frame optical H$\alpha$ emission line. 
Kinematic modelling of both lines concluded the bulk of the molecular gas in ALESS67.1 could be described as an on-going merger, although without being able to reject a rotating disk given the errors.
It is still uncertain whether ALESS67.1 is a multi-component system in an early state of merging.
This scenario has been discussed in Section \ref{subsec:COline} based on our CO and \textit{HST} observations, and it has also been supported by the analysis of its kinematics based on ancillary data by \citet{chen17}.
For further dynamical mass calculations we will only consider the properties of the most luminous component of ALESS67.1 (Table \ref{tab:COprops}), and assume that this component is well-described by a rotating disk.

ALESS122.1 is the source of highest SNR in our sample and the velocity structure consistent with disk rotation.
Due to these reasons it was possible to analyse its velocity field quantitatively using a rotating disk model, as will be described in Section \ref{subsubsec:galpak}.

Although the velocity gradients may be consistent with large-scale disk rotation as a bulk motion, we emphasize that no conclusions can be drawn to exclude complex, disturbed gas motions on scales smaller than the resolution limit of our data.
Based on this qualitative analysis of the PV-diagrams in Figure \ref{fig:channelmaps}, we will adopt the scenario of a rotating disk for our sources for the computation of their dynamical masses. 
The velocity line width (and thus an estimate of the velocity dispersion) and other morphological properties will be adopted from the values listed in Table \ref{tab:COprops}.

\begin{figure*}
\centering
\includegraphics[trim={1cm 0.5cm 0.5cm 0.5cm},clip,width=0.9\linewidth]{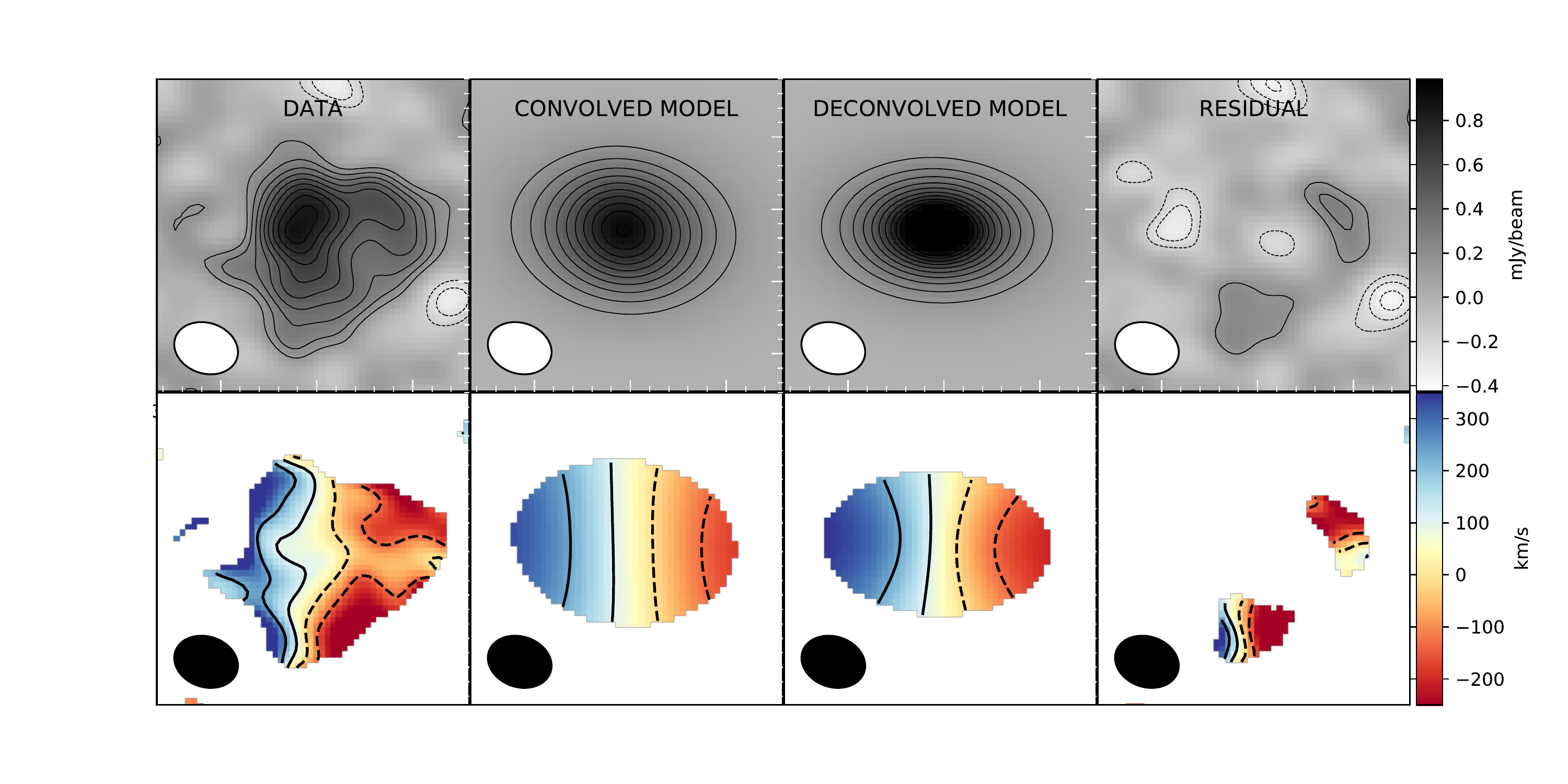}
\caption{ Modelling of the morphology and kinematics of ALESS122.1 with \textsc{GalPak3D}. The top row shows the zeroth-moment maps (integrated intensity) of the observed, convolved model, deconvolved model and residual cubes. The contour lines in the top row show the 2-9 $\sigma_{\rm rms}$ levels, where $\sigma_{\rm rms}$ is the noise level of the zeroth-moment maps.  The bottom row shows the first-moment maps (intensity-weighted velocity) of the corresponding cubes, showing only regions with fluxes $>2 \sigma_{\rm rms}$. The color scale represents the width of the CO line in km s$^{-1}$. The low-intensity levels of the residual maps show that the bulk of the emission can be described by disk rotation. This, however, does not preclude the galaxy from being a merger, as an ordered rotating disk can be quickly 'reformed' after the final coalescence stage in a few dynamical times \citep{robertson06, hopkins13} }
  \label{fig:galpak122}
\end{figure*}
\subsubsection{Kinematic Modelling}\label{subsubsec:galpak}

To quantitatively describe the kinematics of the molecular gas in our sources, we model the data using the modelling package \textsc{GalPak3D}. 
\textsc{GalPak3D} is a Bayesian parametric Markov Chain Monte Carlo fitter for three-dimensional (3D) galaxy data that attempts to disentangle the galaxy kinematics from resolution effects \citep{bouche15}.
Starting from a set of uniform priors on the disk parameters (source center, radius, inclination, velocity dispersion, etc.), a 3D disk galaxy model is produced and compared to the data to compute a reduced $\chi^2_{\nu}$ value, which is then minimized during the sampling process.
The reliability of the inferred kinematic parameters in \textsc{GalPak3D} goes approximately as $\frac{\delta p}{p} \propto \frac{r_{1/2}}{r_{beam}} \times SB_{1/2, obs}$, i.e. the uncertainty in the inference of the parameter $p$ is inversely proportional to the source size to resolution ratio ($\frac{r_{1/2}}{r_{beam}}$) and surface brightness of the source ($SB_{1/2, obs}$).
Given the dependence of these uncertainties on the quality of the data and due to the SNR limits in most of our sources, we were able to apply this method only to the source with highest SNR and resolution, ALESS122.1.

Figure \ref{fig:galpak122} shows zeroth-moment (integrated intensity) and first-moment (intensity-weighted velocity) maps for ALESS122.1.
The four images  
correspond to the moment maps of the observed data cube, the convolved model, deconvolved model and residual (model-substracted) cubes, where the latter three cubes are output products of \textsc{GalPak3D}.
We estimated the noise per pixel in the zeroth-moment map as $\sqrt{N} \times \sigma_{chan}$, where N is the number of channels used and $\sigma_{chan}$ is the rms noise per channel.
Using this value, for the computation of the first moment maps of the four cubes, we masked out all pixels with SNR<2.
The bottom row of Figure \ref{fig:galpak122} shows the first-moment maps (intensity-weighted velocity) of the corresponding cubes. 
The color scale has been chosen to represent the velocity width of CO line of the source.

The resulting 3D model cube shows a modest agreement with the data, with a reduced $\chi^2_{\nu}$ of 1.87, which is calculated over the total area and frequency range covered by the observed and modelled data cubes, as represented in Figure \ref{fig:galpak122}. The slightly-high reduced $\chi^2_{\nu}$ value can be explained by residual structure, which can be seen in the 0th and 1st moment maps of the residual cubes.
The model parameters obtained for the source agree with comparable properties inferred directly from the line profile.

The bulk of the velocity field of ALESS122.1 is consistent with a rotation-dominated disk (mindful of the residual clumps), with an inclination of $i \sim 52^{\circ}$, a maximum rotational velocity of $v_{\rm max} \sim 560$\kms, and a velocity dispersion of $\sigma_v \sim 130$\kms. 
The residual image in the right upper panel of Figure \ref{fig:galpak122} reveals few residual clumps between 2-3 $\sigma$ significance, consistent with being noise.
We conclude that the bulk of the velocity field is to a first order consistent with disk rotation.

\subsection{Dynamical Masses}\label{subsec:dynamicalmasses}

The kinematic properties of a galaxy, obtained e.g. through modelling its velocity field, can provide a reliable estimation of the mass enclosed within the region covered by the emitting medium.
At high redshift, this is a complicated task, since observations of molecular gas are frequently poorly spatially resolved and the morphology of the mass distributions are thus usually unknown.
Based on the kinematic study above and the size estimates from our analysis, we calculate the dynamical masses assuming the bulk of the emission in our sources can be well described by a rotating disk. 
The total dynamical mass within a radius $r=2r_{1/2}$ is then given by:
\begin{equation}
M_{\rm dyn}(r<2r_{1/2}) = \dfrac{(\Delta v_{\rm rot}/2 \sin{(i)})^2 \times 2 r_{1/2}}{G},
\label{eq:mdyn}
\end{equation}
where $r_{1/2}$ is the half light radius estimated through $uv$-fitting measured in kpc, $i$ is the inclination of the galaxy (\ref{tab:COprops}) and $G$ the gravitational constant \citep[e.g., ][]{solomon05, erb06, deblok14}.
In the cases where the velocity field cannot be modelled, the rotational velocity $v_{\rm rot}$ can be estimated as half the velocity width $\Delta V_{\rm FWHM}$ of the line profile. For ALESS122.1 we use the parameter values for $v_{\rm max}$ resulting from the kinematic modelling in Section \ref{subsubsec:galpak}, although both estimates provide consistent results within errors.

A large uncertainty in the estimation of the dynamical masses is contributed by the inclination angle of the disk.
In previous dynamical-mass studies, it has been customary to adopt a
value of $<\sin^2(i)> = 2/3$ (corresponding to $i$=54.7~deg), which is the average value expected for a randomly oriented population of disk galaxies.
However, given our high resolution data, we have enough information to use a simple assumption motivated by the early observation that the apparent axis ratio is closely related to the inclination angle for a disk. 
We use the relation $cos^2(i) = ((b/a)^2-q_0^2)(1-q^2_0)^{-1}$ \citep{hubble26}, where $(b/a)$ is the axis ratio and $q_0$ is the inherent thickness of the disk, assuming $q_0=0.1$ \citep{nedyalkov93} . 
Obtaining the deconvolved axis ratios using the CASA routine \textsc{imfit}, we compute the inclination angles for ALESS49.1, ALESS57.1, ALESS67.1 and ALESS122.1 to be $i\sim 80\pm30^{\circ}, 40\pm50^{\circ}$, $50\pm20^{\circ}$ and $40\pm20^{\circ}$, respectively.
For ALESS122.1, where a more complex analysis was possible in Section \ref{subsec:kinematics}, we will adopt the inclination angle estimated through the kinematic modelling, $i\sim 52\pm2 ^{\circ}$, which is in agreement with the axis-ratio approximation within the errors. 
Although this is an approximation and the uncertainties are large, this estimation is superior to assuming the single value of $<\sin^2(i)> = 2/3$ for all systems, since the inclination in individual sources may anti-correlate with line width. 
It is interesting to note the high inclination angle of ALESS49.1, which is also supported by the small CO size and double-peaked line profile.
Although these profiles are not uncommon for rotating disks and have been found in at least 40\% of SMGs \citep{tacconi08, bothwell13}, the asymmetry between the double-peaked profile suggests a non-uniform distribution of the gas in the 'disk', produced possibly by minor instabilities or an unresolved merger of two gas disks. 

The resulting dynamical masses for our sample range from 1--5$\times 10^{11}$\msun (Table \ref{tab:COprops}), as calculated within 2$\times$ the half-light radii of the sources $r_{1/2}$. 
These values are in general agreement with previous measurements of the dynamical masses of two of these sources based on other tracers \citep[CO($J$=1-0) and H$\alpha$,][]{huynh17, chen17}.
\footnote{In a study of the CO(1-0) emission, \citet{huynh17} found equivalent dynamical masses of $M_{\rm dyn} \sin{i}^{2} = (2.1 \pm 1.1)$ and $(3.2 \pm 0.9) \times 10^{11} M_{\odot}$ for ALESS122.1 and ALESS67.1, respectively.
Although discrepancies were expected given that they used optical instead of CO extensions due to the low resolution of their CO data and assumed an inclination angle of $\sin^2(i)=2/3$, their result is consistent with ours given the errors.
Similarly, a dynamical study of the $H\alpha$ emission in ALESS67.1 presented by \citet{chen17} estimated a dynamical gas mass of $ M_{\rm dyn}= (2.2 \pm 0.6) \times 10^{11} M_{\odot}$ for ALESS67.1.
Although this estimation resulted from the analysis of the total source, while our calculation used only the main component of the source, the values estimated agree with our calculations within the errors.}
This range corresponds to masses at the high end of the average values found for other SMG samples by \citet{ivison10} (2.3$\times 10^{11}$M$_{\odot}$), \citet{tacconi08} (1.3$\times 10^{11}$\msun), \citet{engel10} ($ \sim 1.9 \times 10^{11}$\msun) and slightly below the values found for some extreme sources such as GN20 \citep[$5.4\pm 2.4 \times 10^{11}$\msun,][]{hodge12} and SMMJ131201 \citep[$9.5\pm 2.4 \times 10^{11}$\msun ,][]{engel10}.
The uncertainties in these values are propagated from all parameters used in their calculation, including the inclination angles $i$ (Table \ref{tab:COprops}).

\subsection{Implications on $\alpha_{CO}$ and $M_{*}/L$ estimates}\label{subsec:covariance}

\begin{figure}
     \includegraphics[trim={0.51cm  0.4cm 1.9cm 0.2cm},clip,width=1.04\linewidth]{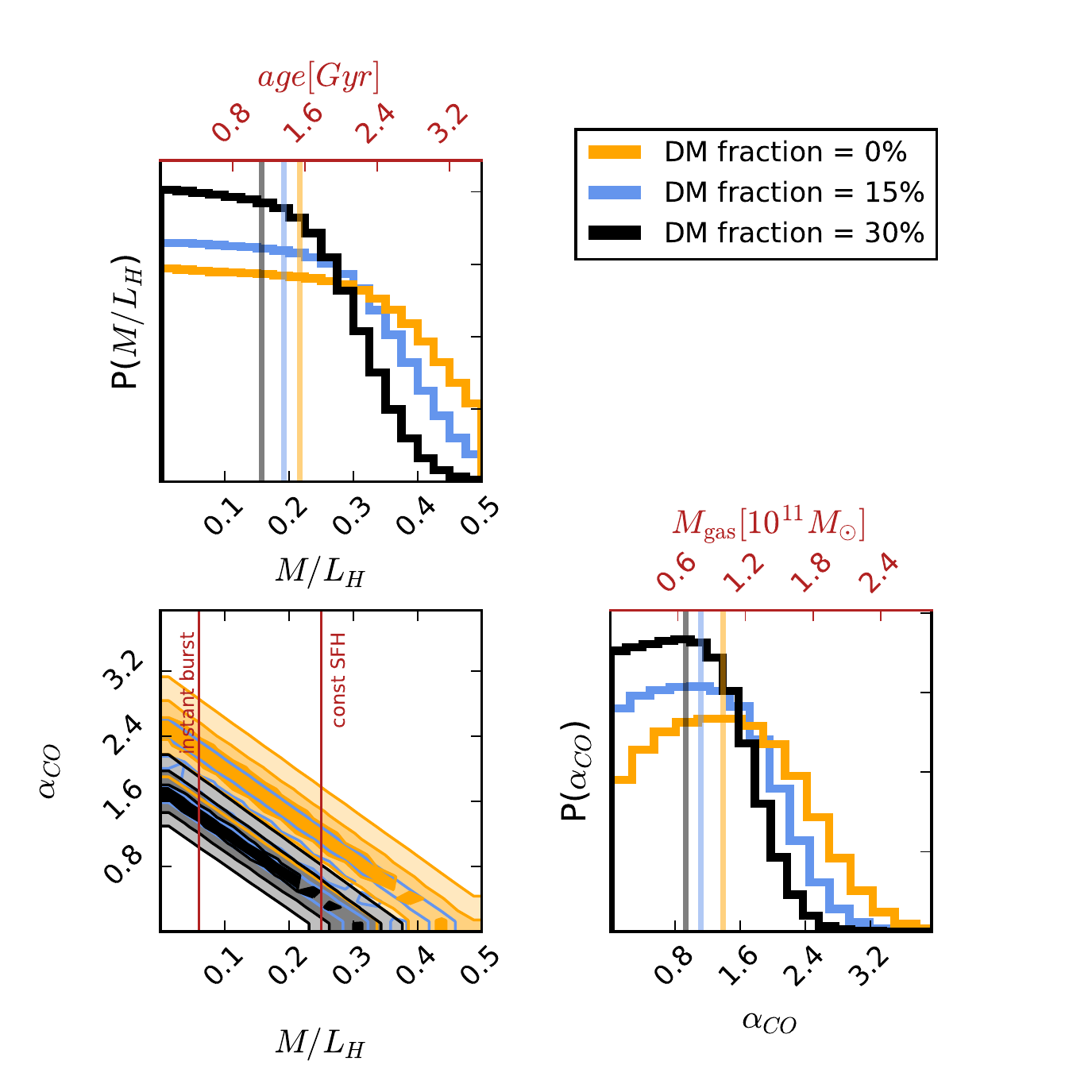}
    
    \caption{One- and two-dimensional posterior probability density functions (PDFs) of the $M_{*}/L_{H}$ and $\alpha_{\rm CO}$ parameters for the ALESS sources in this work.
    \textit{The upper and lower-right panels} show the one-dimensional PDFs as histograms, where the orange, blue and black solid lines correspond to the inference assuming dark matter contributions of 0\%, 15\% and 30\%, respectively. The galaxy age axis (upper panel, in red) equivalent to the $M_{*}/L_{H}$ axis is computed using the relation presented by \citet{hainline11} for an instantaneous burst SFH. The $M_{\rm gas}$ axis (lower-right panel, in red) is computed by adopting the median $L_{\rm CO}$ of our sample, $<L_{\rm CO}>= 7.7 \times 10^{10}$ K km/s pc$^{2}$. The transparent orange, blue and black vertical lines correspond to the median values of the distributions. \textit{The lower-left panel} shows the covariance plot (two-dimensional PDF) of the parameters, were a clear correlation can be recognized through the diagonal shape of the contours, which represent the 25th, 50th and 75th percentiles of the sampled distribution.Two reference lines (red) are drawn, corresponding to the average $M_{*}/L_{H}$ ratio found by \citet{simpson14} for the ALESS survey as a whole, assuming a constant SFH in contrast to an instantaneous burst SFH.  }
    \label{fig:covariance}
\end{figure}

The dynamical mass estimates discussed in Section \ref{subsec:dynamicalmasses}
can be related to the various mass components in galaxies following:
\begin{equation}
M_{dyn}(r \leq 2 r_{1/2}) = M_{\rm baryons}(r \leq 2r_{1/2}) +M_{\rm DM}(r \leq 2r_{1/2}),
\label{eq:Mtot}
\end{equation}
where $ M_{\rm baryons}=M_{\rm gas} +M_{*}$ and $M_{\rm DM}$ is the dark matter contribution.
Thus, given assumptions on the dark matter (DM) content and stellar masses, the dynamical mass estimates can be used as an independent method to constrain the total gas masses in galaxies.
Before we proceed to constrain the gas masses using this method, we summarize the unknown parameters intrinsic to this calculation.
We note that given the scarcity of CO($J\leq3$) gas data at high resolution, most of the existing literature on molecular gas in high-redshift galaxies usually assumes fixed values for unknown parameters, estimating gas masses which are thus sensitive to these assumptions and lacking information on the inherent systematic uncertainties.

To begin with, the H$_2$ molecules that comprise the bulk of the molecular gas reservoir in galaxies are not directly observable. 
Measurements of molecular gas thus rely on CO observations, via a conversion from the ground-state CO(1-0) luminosity which is parametrized with the factor $\alpha_{\rm CO}$. 
Indeed, high-redshift measurements are providing increasing evidence that the $\alpha_{\rm CO}$ values in the early universe can be lower than in solar-metallicity galactic disks \citep[$\alpha_{\rm CO}\sim0.8$--1.0, e.g., ][]{tacconi08, hodge12, bothwell13}.

In addition to the uncertainties in the total gas mass estimation, stellar mass estimates are typically obtained via SED-fitting, which relies on assumptions regarding the star formation history (SFH) of the stellar populations and dust attenuation, which can be uncertain especially for starburst systems (such as SMGs).
In particular, it has been increasingly observed that SED fitting of starburst systems suffers from a degeneracy between the SFH, the mass-to-light ratio $(M_{*}/L_{\rm H})$ and the age of the galaxy \citep{hainline11, simpson14}.
This degeneracy means that in a galaxy with an instantaneous-burst SFH the total stellar mass would be contained almost exclusively within young luminous stellar populations, resulting in a low $(M_{*}/L_{\rm H})\sim 0.05$. 
Conversely, in a galaxy of intermediate age with a constant SFH, most of the stellar mass
would be distributed in old faint populations whose emission is outshone by the luminous newly born stars, 
implying a higher mass-to-light ratio ($M_{*}/L_{\rm H}\sim 0.3)$.
As shown by recent resolved SED-fiting studies \citep{sorba18}, this effect could underestimate total stellar masses by factors of up to 5, especially in sources of high specific star-formation rates such as SMGs.

Rather than adopting a single $(M_{*}/L_{H})$ or $\alpha_{\rm CO}$, we can instead parametrize the baryonic mass as:
\begin{equation}
M_{\rm baryons} = L_{H}\times M_{*}/L_{H} + \alpha_{\rm CO} \times L_{\rm CO}, 
\label{eq:Mbaryons}
\end{equation}
where $L_{H}$ is the rest-frame $H$-band luminosity corrected for dust obscuration
(here estimated by \citealt{dacunha15}) and $L_{\rm CO}$ is the CO($J$=1-0) luminosity estimated in this work.

Finally, the dynamical mass of a galaxy also includes the mass of the non-baryonic dark-matter (DM) component, which is a source of uncertainty, as no independent measurement of this mass fraction is available for our SMGs.
Unlike observations in local disk galaxies, which have reported dark matter fractions ($f_{\rm DM}$) of $\sim 50$\% \citep[e.g., ][]{courteau15}, it has been claimed by recent spectroscopic surveys \citep[e.g., ][]{price16, wuyts16, genzel17} that more compact star-forming disk galaxies at $z>2$ appear to be heavily baryon dominated with $f_{\rm DM}\sim$10--20\%, although these calculations involve the same unknown factors as are used for the gas fractions.
In contrast, recent simulations \citep[e.g.][]{lovell18} have reported that the dark matter fractions in disc-like galaxies could range up to 65\% for galaxies with stellar masses similar to SMGs ($\sim 10^{11}\rm M_{\odot}$).

Combining equations \ref{eq:Mtot} and \ref{eq:Mbaryons}, the total mass can be expressed as:
\begin{equation}
M_{\rm dyn} = \dfrac{L_{H}\times (M_{*}/L_{H}) +  L_{\rm CO} \times \alpha_{\rm CO} }{ (1-f_{\rm DM})}, 
\label{eq:inference}
\end{equation}
where the conversion factor $\alpha_{\rm CO}$, the stellar mass-to-light ratio $(M_{*}/L_{H})$ and the dark matter fraction $f_{\rm DM}$ are the unknown parameters. Given the small number of independent data points available (one for each of our four sources), in order to reduce the parameter space we will keep the $f_{\rm DM}$ value fixed, and discuss the effect of low and high $f_{\rm DM}$ values to the other parameters distributions later on.

Firstly, we reproduce the parameter space built up in Equation \ref{eq:inference} by applying an MCMC technique, using the open-source algorithm \textsc{emcee} \citep{fm13} for the sampling.
Based on the likelihood of the measured dynamical masses $M_{\rm dyn}$ (estimated in Section \ref{subsec:dynamicalmasses} as a function of $L_{\rm CO}$ and $L_{\rm H}$) given this model, we sample the posterior probability density function (posterior PDF) for $\alpha_{\rm CO}$ and $(M_{*}/L_{H})$. 
It is important to note that at this point the likelihood function accounts only for the uncertainties in the dynamical masses $M_{\rm dyn}$.
To account for the uncertainties in the $L_{\rm CO}$ and $L_{\rm H}$ variables as well, we use a Monte Carlo approach and repeat the inference exercise presented above for 100 iterations, each time using values randomly chosen from within a normal distribution that corresponds to the mean and standard deviation of our $L_{\rm CO}$ and $L_{\rm H}$ measurements.
Finally, we average the posterior PDFs gathered in these 100 iterations and obtain a final probability distribution for the parameters which takes into account the uncertainties in the measurements of all observables ($M_{\rm dyn}$, $L_{\rm CO}$ and $L_{\rm H}$).

Fig. \ref{fig:covariance} shows the one- and two-dimensional final posterior PDFs of the $(M_{*}/L_{H})$ and $\alpha_{\rm CO}$ parameters adopting three different dark-matter fractions $f_{\rm DM}= 0\%, 15\%$ and 30\%.
The covariance between the parameters can be recognized in the two-dimensional PDF (Figure \ref{fig:covariance}) as an elongated ellipse in the central countour level plot.
Higher DM fractions decrease the value of $\alpha_{\rm CO}$ and consequently $M_{\rm gas}$. This trend is similarly followed by the $(M_{*}/L_{\rm H})$ parameter.
For a DM fraction of 15\% we find a median value of $\alpha_{\rm CO}= 1.1^{+0.8}_{-0.7}~[\rm M_{\odot}/ (K~km~s^{-1}~pc^{2})^{-1}]$, while for a larger DM fraction of 30\%, these values decrease to $\alpha_{\rm CO}= 0.9^{+0.7}_{-0.6}~[\rm M_{\odot}/(K~km~s^{-1}~pc^{2})^{-1}]$. 
For $f_{\rm DM}= 0$ we obtain an upper limit to $\alpha_{\rm CO}\leq  1.4^{+0.9}_{-0.9}~[\rm M_{\odot}/ (K~km~s^{-1}~pc^{2})^{-1}]$.
For the mass-to-light ratio parameter we find  $(M_{*}/L_{\rm H})= 0.22^{+0.16}_{-0.15} \rm M_{\odot}/L_{\odot}$, $(M_{*}/L_{\rm H})= 0.19^{+0.15}_{-0.13} \rm M_{\odot}/L_{\odot}$ and $(M_{*}/L_{\rm H})= 0.16^{+0.12}_{-0.10} \rm M_{\odot}/L_{\odot}$ for DM fractions of 0\%, 15\% and 30\%. 
Using these values, the gas reservoirs in our sources have an average gas mass of $\sim 0.9 \times 10^{11}$\msun (DM contribution of 15\%) and $\sim 0.7 \times 10^{11}$\msun~ (DM contribution of 30\%), corresponding to average gas-to-total-mass fractions of  0.26\% and 0.33\%, respectively.
These sources show similar gas fractions than those reported for typical star-forming galaxies \citep[$\leq 50$\%, ][]{tacconi17}. 
For comparison, \citet{bothwell13} presented a study of a representative sample of SMGs at $z\sim$2 from predominantly CO(3-2) observations, finding mean gas masses of $(3.2 \pm 2.1) \times 10^{10}$\msun assuming $\alpha_{CO}=1$. 
Since the resulting PDF for the $M_{*}/L_{\rm H}$ and $age$ parameters for our galaxies is broad and poorly constrained, the galaxy stellar population properties inferred through this method ($\sim 0.1-2.0$ Gyr) easily agree with those estimated through SED-fitting by \citet{dacunha15}, which are also associated with large errors ($\sim$70\%). 
They find the age of these galaxies range between $\sim$ 0.04--1.22 Gyr, corresponding to $M_{*}/L_{\rm H}$ between 0.16--0.2.

As discussed above, the values recovered for the parameters are subject to large uncertainties due to our small sample size.  
However, we note that one goal of this exercise is to shed light upon the degeneracies among the key parameters and large uncertainties inherent to gas mass calculations.
Most importantly, given the increasing availability of dynamical mass measurements for high-redshift galaxies with ALMA, we consider that this approach has potential for constraining the molecular gas properties with more accuracy, especially since larger sample sizes will allow the exploration of a more detailed parameter space in the future.


\section{Distributions of molecular gas, dust continuum and stellar emission}\label{sec:distributions}

\subsection{Offset distributions of ISM tracers and stellar emission in ALESS122.1}
\label{subsec:offset}


\begin{figure}
\centering
 	 \includegraphics[trim={2.4cm  0cm 1cm 0.2cm},clip,width=1.\linewidth]{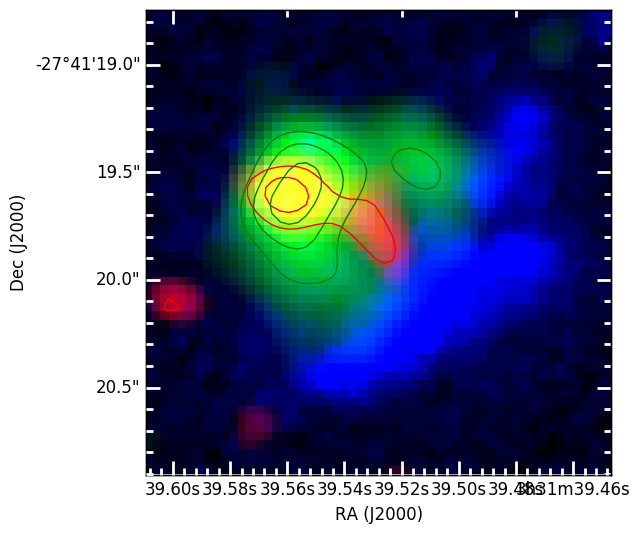}
 	 
    \caption{False-color image of the molecular gas, dust continuum and stellar emission components in ALESS122.1. The molecular gas (green) shows the CO(3-2) emission and the dust continuum region (red) corresponds to the 3 mm (rest-frame $\sim 850~\mu$m) continuum detection (4-$\sigma$), both masked at $S \geq 3\sigma$. For clarity we overplot contours for both the CO ($6,8,10\sigma$) and dust continuum emission ($3,4\sigma$), to show that their centroids are co-located. The stellar emission (\textit{HST}-ACS F814W) is presented as the blue region and shows a large offset from the gas and dust centroid, with almost no overlap even when considering the combined astrometric uncertainties. The \textit{HST} astrometry in this data is accurate since the images were corrected based on \textit{GAIA} measurements. The increasing evidence of a population of sources with misaligned stellar, dust continuum and gas emission has important implications for energy balance assumptions in SED-fitting. We discuss the striking difference between the observed extent of the gas and dust continuum emission for a larger sample of sources in Section \ref{subsec:stackedprofiles}.}
    \label{fig:rgb}
\end{figure}
An interesting case among our sample is the apparently uncorrelated distribution of dust continuum, gas and stars in ALESS122.1 (Fig.~\ref{fig:rgb}).
The gas and stellar emission extend across regions of similar size, separated by $\sim$ 0.5$''$ ($\sim$ 5~kpc at $z = 2$) as measured from their centroids, but are tightly aligned next to each other.
There is almost no overlap of gas and stellar emission, even when considering the combined astrometric uncertainties ($<$0.1$''$\footnote{The \textit{HST} astrometric uncertainties are low since the image has been accurately calibrated using \textit{GAIA} measurements of this field.}). 
Moreover, while the gas and stars have half-light radii $r_{1/2}\sim4$~kpc and are thus extended on scales of $\sim$10~kpc, the detected dust continuum emission (3mm observed-frame, corresponding to $\sim 850~\mu$m rest-frame) is confined to a central region of $\sim$5~kpc, with the emission peak spatially coincident with that of the molecular gas emission.
Similar physical offsets between dust continuum, gas and stellar emission, and specifically between ALMA and \textit{HST}, have been previously found in a number of high-redshift sources \citep[e.g., ][]{riechers10, chen15, hodge15, fujimoto17, simpson17}. 
For example, based on low-resolution 870$\mu$m continuum data ($\sim$1.5$''$),
\citet{chen15} reported a statistical offset
with the existing stellar component traced by \textit{HST} 
as large as $\Delta p$ = 0.4$''$ ($\sim$ 4 kpc at $z = 2$) in  66\% of the ALESS SMGs.
Similarly, \citet{hodge15} found in the $z=4$ galaxy, GN20, that the FIR and CO emission are offset by 0.6$''$ (4 kpc) from the peak of the rest-frame UV emission as traced by the \textit{HST}/WFC3 F105W image.

We consider three scenarios for explaining the mismatch of the distributions of stellar and gas/dust continuum emission in ALESS122.1.
The first and most plausible scenario assumes that the optical component extends to the regions of gas and dust continuum emission, while suffering extreme extinction in these regions.
Such high-extinction scenarios have been suggested previously in starbursts with extreme SFRs and in cases of partial or total absence of optical counterparts \citep[e.g., ][]{walter12, hodge12, simpson15}.
This is not uncommon in sub-mm selected samples, where optical detection rates are only $\sim 70-80$\% even with deep data \citep[e.g., ][]{simpson14}. 

The second scenario implies that the observed offset reflects a physical misalignment between the gas/dust continuum distributions and that of the stellar mass.
This argument has been tested by \citet{chen15} for the full ALESS sample. 
They tested this hypothesis by comparing the positional offsets found in low-redshift ($z<2$) to those in high redshift ($z>2$) subsamples, taking into account that at lower redshift the optical photometry would probe wavelengths less affected by obscuration. They found no statistical difference in the measured positional scatter between the low- and high-redshift samples, implying that obscuration is not the dominant factor for the population as a whole. 

The third scenario implies that ALESS122.1 is an ongoing, merging system comprised of two components, one of which is a heavily obscured SMG which has no detected optical counterpart, and the other is an optically bright galaxy in the \textit{HST}-ACS  F814W imaging (which corresponds to rest-frame $\sim$ 3000 \AA).
High-resolution optical-MIR counterparts at other wavelengths are needed in order to confirm the nature of this source and explain the uncorrelated distributions of its physical components. 

Regardless of the underlying scenario, the significant anticorrelation between dust continuum and optical imaging disprove some assumptions which are intrinsic to multiwavelength studies such as energy balance in SED-fitting \citep[this has been previously discussed by e.g., ][]{hainline11, chen15, hodge16, simpson17, chen17}.


\subsection{Statistical analysis on the relative sizes of dust continuum, molecular gas and stellar emission in SMGs}\label{subsec:stackedprofiles}

To gain a general understanding of the distributions of the molecular gas, dust continuum, and stellar emission in SMGs and how they relate, next we conduct a stacking analysis using the CO emission measured in our sources and the available multiwavelength data at similar resolution.  

\begin{figure*}
\centering
 	 \includegraphics[trim={1.9cm  0cm 1cm 0.2cm},clip, width=1.0\linewidth]{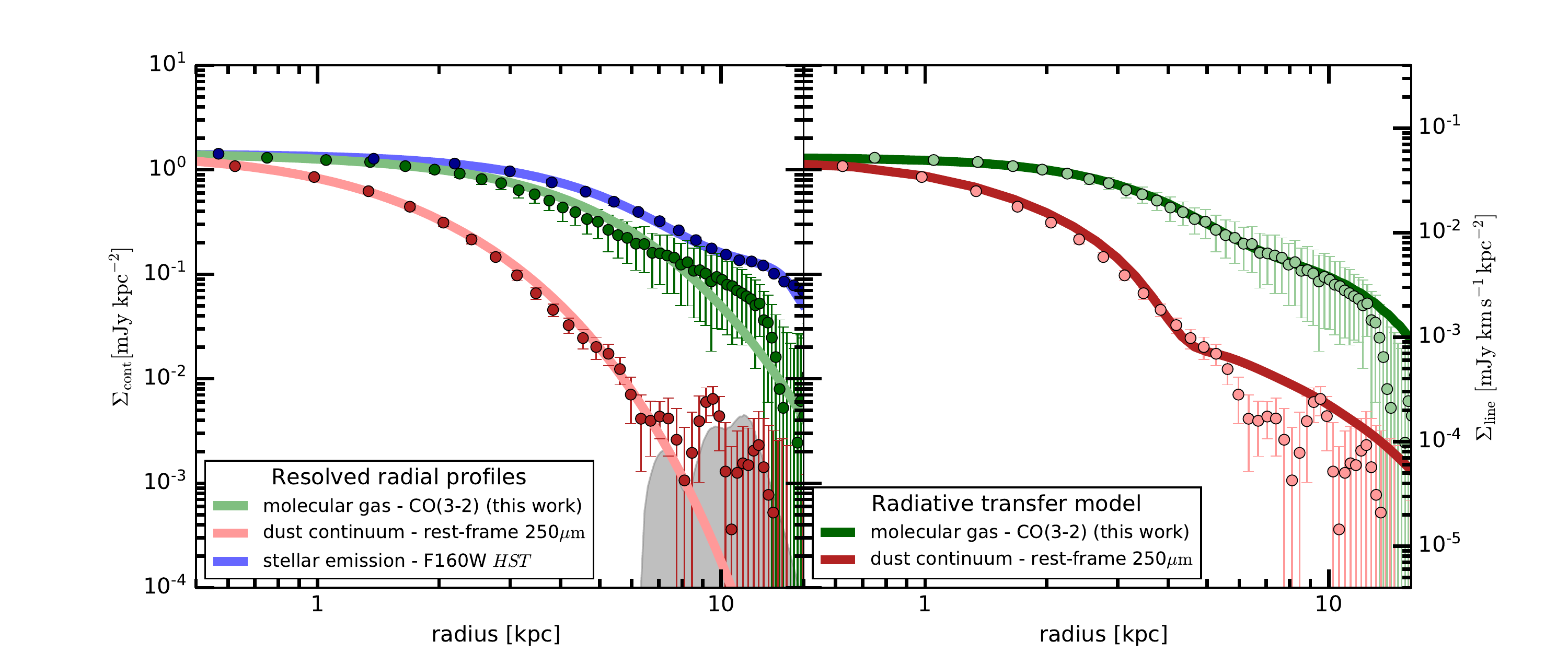}
    \caption{Stacked radial profiles and modelling. \textit{Left panel:} one-component exponential profile fits are shown for a) the stacked CO(3-2) emission presented in this paper (green solid line), b) the stacked high-resolution dust continuum emission at 870 $\mu$m (rest-frame $\sim 250~\mu$m) of 16 luminous ALESS SMGs by \citet{hodge16} (red solid line) and c) the stacked high-resolution H-band emission (\textit{HST}) for the same 16 ALESS sources presented by \citet{hodge16} (blue solid line). The profiles were normalized to have the same peak flux densities, and the error bars were obtained through bootstrapping. All exponential fits are convolved with Gaussian functions of the size of the beam, to account for the effect of different resolutions. The data points of the dust profile at radii of $>$7 kpc were discarded for the fitting, since these are evidently affected by artifacts produced by the side lobes of the beam (shown as the grey shaded area). We find that the cool molecular gas and stellar emission are clearly more extended than the rest-frame 250 $\mu$m dust continuum by a factor of $>2$. \textit{Right panel:} To explain the different sizes, we fit an extended version of the dust and line radiative transfer models of \citet[][]{weiss07} to our data.  The solid lines are the profiles resulting from the best-fit radiative transfer model. The model assumes an exponentially decreasing column density distribution (S\'ersic index of $n=1$) and a linear temperature gradient decreasing from the center to the outer parts of the disk. Solving the dust and line radiative transfer in several radial bins, we obtain a exponentially decreasing gas and dust radial profile, taking into account the different beam sizes for the dust continuum and CO line observations. Assuming a constant dust-to-gas ratio, the model is thus able to reproduce the apparent size differences between the CO(3-2) and dust continuum emission by only introducing temperature and optical depth gradients.}
    \label{fig:stacks}
\end{figure*}

We first produced the stacked radial profiles of the CO emission of the four ALESS sources presented here.
To achieve this, we create $10\times10''$ thumbnails of our sources, align their centers and scale them to have the same peak flux. 
We then produce the radial profiles of the individual sources by radially averaging the emission centered at the source peak, and finally, we average the individual radial profiles in our sample (Figure \ref{fig:stacks}).\footnote{We note that we investigated the difference between the approach described above and an approach in which a stacked image is produced first and the mean radial profile is then extracted directly from the stacked image. We find that both methods are equivalent and result in identical average radial profiles.}
To account for the effect of outlier structures of the individual sources, we estimate the uncertainties of the radial profiles through bootstrapping, e.g., by creating 1000 different average profiles, where each iteration used a randomly resampled combination of sources.

Using the same technique, we also stack high-resolution dust continuum emission (0.16$''$ FWHM at rest-frame 250 $\mu$m) of 16 luminous ALESS sources studied by \citet{hodge16}.
This sample of sources are representative of the ALESS survey as a whole \citep[ALESS median $z = 2.5 \pm 0.2$ and $L_{\rm IR} = (3.0\pm 0.3)\times 10^{12} L_{\odot}$,][]{simpson14, swinbank14} as they have a median redshift of $z = 2.6 \pm 0.5$ and infrared luminosity of $L_{\rm IR} = (3.6 \pm 0.9) \times 10^{12} L_{\odot}$.
Most importantly, these properties are also comparable to those of the sample used for the CO stacked profiles  (median redshift of $z = 2.5 \pm 0.4$ and infrared luminosity of $L_{\rm IR} = (5.6 \pm 2.2) \times 10^{12} L_{\odot}$), although both samples have only one source in common (ALESS67.1).

Finally, we stack high resolution \textit{HST} imaging (0.17$''$ in the $H_{160}$ band) available for the same 16 ALESS sources following the procedure described above.
The stacked surface brightness profiles for the gas, dust continuum and stellar components are shown in Fig. \ref{fig:stacks} as green, red and blue data points, respectively.

To quantify the extent and distribution of the molecular, sub-mm and optical/near-infrared emission, first we assume that the observed emission can be described by a exponential profile (S\'ersic profile with index $n$=1) convolved with a Gaussian profile, which corrects for the effect of the shape of the beam.
Based on this we fit a one-dimensional exponential to the CO and dust continuum profiles and find that they are well described by such a model.
We note that for fitting the dust profile we discarded the data points at radius of $>$7 kpc, since these are evidently affected by artifacts produced by the side lobes of the beam (shown as the shaded area in Fig. \ref{fig:stacks})\footnote{This issue is not relevant for CO, since the side lobes appear only at r>15 kpc}.
The radial profile of the stellar emission results in a bad quality fit when using an exponential, i.e. when fixing the S\'ersic index $n=1$. 
To improve this, we leave the S\'ersic index as a free parameter and achieve a better quality fit with $n=0.8\pm0.2$, which is consistent within the errors with the median value found for the ALESS sample as whole \citep[$n=1.2\pm0.3$, ][]{chen15}.
The best-fit models consist then of exponential profiles with radii of $r= 1.7\pm0.1, 3.8\pm0.1$ and $4.0\pm2.0$ kpc for the dust continuum, CO and stellar emission respectively.

These results reveal that both the stellar ($H_{160}$) and molecular gas (CO($J=3-2$)) emission have similar sizes, but are $>2\times$ more extended than the dust continuum emission, measured at rest-frame 250 $\mu$m.
In addition, we note that the estimated size of the CO(3-2) emission should be considered a lower limit for the extent of the total molecular-mass distribution, as the ground-state  transition CO(1-0) has been observed to be even more extended than the higher $J$-transitions \citep{ivison11, riechers11e}. 

\subsection{Physical implications of the more extended molecular gas distributions as compared to dust continuum emission}\label{subsubsec:stackinterpretation}

A crucial consequence of the different scales probed by the molecular gas, dust and stellar tracers in these galaxies (as shown in Section \ref{subsec:stackedprofiles}) is the uncertainty in using conversion factors between the luminosities of individual tracers.
The compactness of the dust continuum emission in SMGs has been discussed in the literature \citep{simpson15, ikarashi15, hodge16, barro16, simpson17}, as ALMA continuum observations at high angular resolution (0.13--0.4$''$) have revealed median radii of $\sim$ 0.7-- 1.5 kpc.

Apart from CO and optical imaging, high-resolution observations of other SF tracers such as synchrotron and free-free emission have allowed comparison of relative sizes of dust continuum and gas emission regions.
Indeed, radio interferometric observations of SMGs with the Very Large Array
at 1.4 GHz \citep{chapman04, biggs08}, at 3 GHz \citep{miettinen15, miettinen17} and at 6 GHz \citep{thomson18}, have shown that dust continuum sizes appear to be 1.4--4.4 times smaller than the radio.

What is the origin of the significantly smaller sizes ($>2$) of the dust continuum emission in SMGs compared to other physical components?
There are three parameters that could affect the emission of dust at larger radii: the dust-to-gas ratio, the dust temperature and the optical depth of the dust and gas.
The small extent of the dust continuum emission at rest-frame 250 $\mu$m is likely to trace the heating of dust by star formation or AGN, as it is confined to central compact regions, as expected for highly star-forming systems.
A decrease in optical depth at larger radii, would similarly affect the observed sub-mm continuum emission, as would a radially dependent dust-to-gas ratio. Studies in the nearby Universe, however, \citep[e.g.,][]{sandstrom13, groves15} suggest that no large variations in the latter are expected across the galaxy, making this scenario less plausible.

The different spacial extent for the rest-frame 250 $\mu$m dust continuum emission and the CO line emission can be explained in the context of self-consistent radiative transfer of the CO(3-2) and dust continuum emission. 
\citet{weiss07} presented such a model, where the line and dust continuum emission are linked through a constant dust-to-gas mass ratio throughout the galaxy, specifically, by connecting the gas column density $N_{\rm H_2}$ to the optical depth of the dust, $\tau_{\rm dust}(r)\sim N_{\rm dust} \sim N_{\rm H_2}$.
The different dust continuum and CO sizes can then be explained by assuming a radially decreasing gas column density and a radially decreasing temperature distribution $T(r)$.
Following the radiative transfer equation, and neglecting
the background temperature, we have 
\begin{equation}
S(r) \propto T(r) \cdot(1-e^{-\tau}),
\end{equation}
where $T(r)$ implies  $T(r)=T_{\rm dust}(r)$ for the dust and $T(r)=T_{\rm ex}(r)$ for the molecular gas.
We can expect that the long wavelength dust continuum emission, which is optically thin ($\tau\ll0$), will be sensitive to both the column density and the temperature gradients ($S_{\rm cont}(r) \propto T_{\rm dust}(r) \cdot \tau_{\rm dust}(r)$), whereas the optically thick CO line emission ($\tau\gg1$) will to first order only respond to the temperature gradient ($S_{\rm CO}(r) \propto T_{\rm ex}(r)
\sim T_{\rm kin}(r)$ for the low-J CO transitions, which are close to
thermalisation. 
This would have the effect that the dust continuum emission becomes fainter more rapidly as a function of radius than the gas emission.

Based on these arguments, we have used an extended version of the dust and line
radiative transfer models of \citet[][]{weiss07} to demonstrate this
effect for the radial distribution of the stacked CO(3-2) and
rest-frame 250 $\mu$m dust continuum data shown in Figure \ref{fig:stacks}. 
In the model we use an exponentially decreasing column density distribution (S\'ersic index of $n=1$) and a linear temperature gradient from the center to the outer parts of the disk, and we solve the dust and line radiative transfer in
several radial bins (more details on the model will be presented in
Wei\ss~et al. (in prep)). 
Figure \ref{fig:stacks} shows a fit of a exponentially decreasing gas component, taking into account the different beam sizes for the dust continuum and CO line observations.  
Given the uncertainties of the data, the model can reproduce the different radial behaviour of the dust continuum and the CO(3-2) line emission. 
From this fit we derive an intrinsic exponential scale length of 3.4 kpc (0.4$''$) for the underlying H$_2$ distribution, intermediate between the intrinsic sizes we have estimated earlier based on the de-convolved intensity profiles for the dust and the CO(3-2) line emission.

In this section we have proposed that assuming a unique intrinsic distribution of dust and gas, temperature and optical depth gradients alone could give rise to the apparent size differences observed between the CO and dust continuum emission.
If these parameters are the origin of the compactness of dust continuum emission observed in the high-redshift Universe (here probed at rest-frame $\sim250~\mu$m), caution must be exercised when extrapolating properties of high-resolution continuum observations to conclusions on the molecular phase of the ISM.
Specifically, these results advise against combining unresolved molecular gas observations with radius estimates from resolved dust continuum observations when calculating key parameters such as dynamical masses. 
In order to empirically test the assumptions placed into this model on the temperature and optical depth gradients, high-resolution observations of the dust continuum at different frequencies would be necessary, as a gradient in observed sizes of the emission regions is a direct tracer of the dust optical depth across the SED.


\section{Summary}
\begin{itemize}

\item We have presented ALMA observations of the gas and dust of four luminous sub-millimetre galaxies at $z\sim2-3$ to investigate the spatially resolved properties of the inter-stellar medium (ISM) on scales of a few kpc. 
The molecular gas in these sources, traced by the $^{12}$CO($J$=3-2) emission, is extended over FWHM$\sim$5 --14 kpc.
\item We investigated the dynamics of the molecular gas in our sources and modelled the kinematics of one of them, ALESS122.1, finding that the velocity fields of three of our four sources are consistent with disk rotation to first order.
We clarify that this scenario does not preclude the connection of our sources to mergers as observations of nearby merger remnant and simulations show that gas disks reform rapidly within $<$100 Myr of the peak star formation associated with a merger. 

\item The resolved CO imaging provides size measurements that allow us to derive dynamical masses for our sample. 
The dynamical masses found are in the range of (1.1--5.3)$\times10^{11}$\msun, as calculated within $2\times$ the half-light radii of our sources, which is in agreement with other dynamical mass estimates for SMGs.

\item To provide dynamical constraints to the gas masses, we explore the uncertainties introduced by the estimation of stellar masses, the assumptions on the dark matter fractions and the CO-to-H$_{2}$ conversion factor $\alpha_{\rm CO}$.
Taking into account the covariance between the mass-to-light ratio and $\alpha_{\rm CO}$ parameters, we estimate an average CO-to-H$_{2}$ conversion factor of $\alpha_{\rm CO}=1.1^{+0.8}_{-0.7}$ and $\alpha_{\rm CO}=0.9^{+0.7}_{-0.6}$ for dark matter fractions of 15\% and 30\% respectively, and an upper limit of $\alpha_{\rm CO}\leq 1.4^{+0.9}_{-0.9}$ for a dark matter fractions of 0\%.
These values imply gas fractions of $\sim$30\% for our sources, which are similar to those estimated for main-sequence star-forming galaxies and other SMGs.

\item Our high resolution study allows us to investigate the correlation between the spatial distribution of the physical components of the ISM (the dust continuum and gas) and the stellar distributions.
The sizes of gas and stars are comparable but spatially uncorrelated, while the rest-frame 250 $\mu$m dust continuum is significantly more compact.
The observation of the anti-correlated distributions of the dust continuum and gas emission with respect to the unobscured stellar emission may challenge energy balance assumptions in global SED fitting routines and suggests that caution must be exercised particularly for dusty star-forming galaxies, such as SMGs.

\item To investigate this question statistically, we conduct a stacking analysis of available high resolution ancillary data for SMG populations of similar properties.
 We find that the cool molecular gas and stellar emission are clearly more extended than the rest-frame 250 $\mu$m dust continuum by a factor of $>$ 2.

\item We reproduce our observations with a radiative transfer model, finding that the different sizes are consistent with the expected response of optically thin dust and optically thick gas to radially decreasing optical depth and temperature gradients, when a constant dust-to-gas ratio is assumed.
We suggest that  
extrapolations from morphological properties of high-resolution continuum observations to conclusions on the molecular phase of the ISM should be thus treated cautiously.
\end{itemize}

 \section*{Acknowledgements}
 \addcontentsline{toc}{section}{Acknowledgements}
We would like to thank the Allegro group in Leiden, especially Dr. Luke Maud, Dr. Erwin de Blok (Astron) and Dr. Alison Peck for valuable help and advice at different stages of the project.
GCR acknowledge support from the European Research Council under the European Unions Seventh Framework Programme (FP/2007- 2013) /ERC Advanced Grant NEW-CLUSTERS-321271.
JH acknowledges support of the VIDI research programme with project number 639.042.611, which is (partly) financed by the Netherlands Organisation for Scientific Research (NWO).
IRS acknowledges support from the ERC Advanced Grant
DUSTYGAL (321334), a Royal Society/Wolfson Merit Award and STFC (ST/P000541/1).
JLW acknowledges support of an STFC Ernest Rutherford Fellowship (ST/P004784/1) and additional support from STFC (ST/P000541/1).
F.W and B.P.V acknowledge funding through ERC grants "Cosmic Dawn" and "Cosmic Gas".
H.D. acknowledges financial support from the Spanish Ministry of Economy and Competitiveness (MINECO) under the 2014 Ram\'on y Cajal program MINECO RYC-2014-15686.
This paper makes use of the following ALMA data: ADS/JAO.ALMA\#2013.1.00470.S and \#2016.1.00754.S. ALMA is a partnership of ESO (representing its member states), NSF (USA) and NINS (Japan), together with NRC (Canada), MOST and ASIAA (Taiwan), and KASI (Republic of Korea), in cooperation with the Republic of Chile. The Joint ALMA Observatory is operated by ESO, AUI/NRAO and NAOJ.
This research made use of \textsc{astropy}, a community-developed core Python package for Astronomy \citep{astropy}, and the open-source plotting packages for Python \textsc{APLpy} \citep{aplpy}, and \textsc{corner} \citep{corner}.

\bibliographystyle{apj}
\bibliography{ALESS}

\label{lastpage}

\end{document}